\documentclass[structabstract]{aa}

\usepackage{url}
\usepackage[breaklinks=true]{hyperref}
\usepackage{twoopt}
\usepackage[english]{babel}          % English language/hyphenation

\usepackage{natbib}
\bibpunct{(}{)}{;}{a}{}{,} %% natbib format for A&A and ApJ

\usepackage[utf8]{inputenc}
\usepackage{siunitx}
\usepackage{amsmath, amssymb, amsfonts}

\usepackage[farskip=0pt,caption=false]{subfig}

\usepackage{multirow,bigstrut,ctable}
\usepackage[normalem]{ulem}
\usepackage{rotating}
\usepackage[varg]{txfonts} % A&A recommended fonts
\usepackage{threeparttable}
\def \deg         {\text{$^{\circ}$}}
\def \arcmin      {\text{$^\prime$}}
\def \arcsec      {\text{$^{\prime\prime}$}}

\def \mjybeam     {mJy\,beam$^{-1}$}
\def \mujybeam    {$\mathrm{\mu}$Jy\,beam$^{-1}$}

\newcommand{\beam}[2]{{#1}\arcsec$\times${#2}\arcsec}

\def \lol {LoLSS}

\begin{document}

\title{The LOFAR LBA Sky Survey}
\subtitle{II. First data release}
\titlerunning{LOFAR LBA sky survey}

\author{F.~de~Gasperin\inst{1,2}
\and H.~W.~Edler\inst{2}
\and W.~L.~Williams\inst{3}
\and J.~R.~Callingham\inst{3,4}
\and B.~Asabere\inst{4}
\and M.~Br\"uggen\inst{2}
\and G.~Brunetti\inst{1}
\and T.~J.~Dijkema\inst{4}
\and M.~J.~Hardcastle\inst{5}
\and M.~Iacobelli\inst{4}
\and A.~Offringa\inst{4}
\and M.~J.~Norden\inst{4}
\and H.~J.~A.~R\"ottgering\inst{3}
\and T.~Shimwell\inst{3,4}
\and R.~J.~van~Weeren\inst{3}
\and C.~Tasse\inst{6}
\and D.~J.~Bomans\inst{7}
\and A.~Bonafede\inst{8,1}
\and A.~Botteon\inst{8,1}
\and R.~Cassano\inst{1}
\and K.~T.~Chy\.zy\inst{9}
\and V.~Cuciti\inst{2,1}
\and K.~L.~Emig\inst{10}
\and M.~Kadler\inst{11}
\and G.~Miley\inst{3}
\and B.~Mingo\inst{12}
\and M.~S.~S.~L.~Oei\inst{3}
\and I.~Prandoni\inst{1}
\and D.~J.~Schwarz\inst{13}
\and P.~Zarka\inst{14,15}}

\authorrunning{F.~de~Gasperin et al.}
%\offprints{F.~de~Gasperin}

\institute{
INAF - Istituto di Radioastronomia, via P. Gobetti 101, I-40129, Bologna, Italy; \email{fdg@ira.inaf.it}
\and Hamburger Sternwarte, Universit\"at Hamburg, Gojenbergsweg 112, D-21029, Hamburg, Germany
\and Leiden Observatory, Leiden University, P.O.Box 9513, NL-2300 RA, Leiden, The Netherlands
\and ASTRON, the Netherlands Institute for Radio Astronomy, Postbus 2, NL-7990 AA, Dwingeloo, The Netherlands
\and Centre for Astrophysics Research, University of Hertfordshire, College Lane, Hatfield AL10 9AB, UK
\and GEPI, Observatoire de Paris, CNRS, Université Paris Diderot, 5 place Jules Janssen, 92190 Meudon, France
\and Ruhr University Bochum, Faculty of Physics and Astronomy, Astronomical Institute, Universitätsstr. 150, 44801 Bochum, Germany
\and DIFA - Universit\'a di Bologna, via Gobetti 93/2, I-40129 Bologna, Italy, INAF - IRA, Via Gobetti 101, I-40129 Bologna, Italy
\and Astronomical Observatory of the Jagiellonian University, ul. Orla 171, 30-244 Krak\'ow, Poland
\and National Radio Astronomy Observatory, 520 Edgemont Road, Charlottesville, VA 22903, USA
\and Institut f\"ur Theoretische Physik und Astrophysik, Universit\"at W\"urzburg, Emil-Fischer-Str. 31, 97074, W\"urzburg, Germany
\and School of Physical Sciences, The Open University, Walton Hall, Milton Keynes MK7 6AA, UK
\and Fakult\"at f\"ur Physik, Universit\"at Bielefeld, Postfach 100131, 33501 Bielefeld, Germany
\and LESIA, Observatoire de Paris, CNRS, PSL, Sorbonne Univ., Univ. Paris Cit\'e, Meudon, France
\and ORN, Observatoire de Paris, CNRS, PSL, Univ. Orl\'eans, Nan\c{c}ay, France}

%\date{Received ... / Accepted ...}
\abstract
%context
{The Low Frequency Array (LOFAR) is the only existing radio interferometer able to observe at ultra-low frequencies ($<100$~MHz) with high resolution ($<15$\arcsec) and high sensitivity ($<1$~\mjybeam). To exploit these capabilities, the LOFAR Surveys Key Science Project is using the LOFAR Low Band Antenna (LBA) to carry out a sensitive wide-area survey at $41-66$ MHz named the LOFAR LBA Sky Survey (LoLSS).}
%aim
{LoLSS is covering the whole northern sky above declination $24\deg$ with a resolution of 15\arcsec{} and a sensitivity of $1-2$~\mjybeam{} ($1\sigma$) depending on declination, field properties, and observing conditions. Here we present the first data release, including a discussion of the calibration strategy and the properties of the released images and catalogues.}
%methods
{A fully automated pipeline was used to reduce the 95 fields included in this data release. The data reduction procedures developed for this project have general application and are currently being used to process almost all LOFAR LBA interferometric observations. Compared to the preliminary release, direction-dependent errors have been derived and corrected for during the calibration process. This results in a typical sensitivity of 1.55~\mjybeam, four times better than for the preliminary release, at the target resolution of 15\arcsec.}
%results
{The first data release of the LOFAR LBA Sky Survey covers 650~deg$^2$ in the HETDEX spring field. The resultant data products released to the community include mosaic images (I and V Stokes) of the region, and a catalogue of 42\,463 detected sources and related Gaussian components used to describe sources' morphologies. Separate catalogues for the 6 in-band frequencies of 44, 48, 52, 56, 60, and 64 MHz are also released.}
%conclusions
{The first data release of LoLSS shows that, despite the influences of the ionosphere and radio frequency interference, LOFAR can conduct large-scale surveys in the frequency window $42-66$~MHz with unprecedentedly high sensitivity and resolution. The data can be used to derive unique information on the low-frequency spectral properties of many thousands of sources with a wide range of applications in extragalactic and galactic astronomy.}


\maketitle

%---------------------------------------------------------------------
% article body
%---------------------------------------------------------------------
\section{Introduction}
\label{sec:introduction}

% surveys
The LOw Frequency ARray \citep[LOFAR;][]{vanHaarlem2013} is currently the world's largest and most sensitive radio telescope in the radio window covering $10-240$~MHz (wavelength $\lambda = 1-30$ metres). Unlike higher frequency “dish” radio interferometric arrays, the basic antenna elements of LOFAR are wide-field dipoles. There are two separate dipole types at each LOFAR station, covering frequencies above and below 100 MHz. The Low Band Antenna (LBA) operates at $10-90$ MHz and the High Band Antenna (HBA) covers the $120-240$ MHz band. The intermediate region is dominated by radio frequency interference (RFI) due to FM radio stations and is not accessible for observations.

% LBA capabilities
The LOFAR LBA can reach frequencies close to the ionospheric plasma cutoff frequency (typically $\sim 10$~MHz), the ultimate boundary for ground-based radio astronomy. The history of astronomy shows that one of the most effective ways of making fundamental discoveries is to open up new regions of the electromagnetic spectrum to observation. As the only array currently capable of operating below 100\,MHz over baselines of tens to thousands of kilometers, the International LOFAR Telescope (ILT) is a unique instrument for exploring the low radio frequency Universe at high angular resolution. 

% science case
The relativistic electrons observed at LBA frequencies are less energetic and in general trace longer timescales than those observed at higher frequencies. Studying these "fossil", ultra-steep spectrum sources at the high resolution of the ILT can provide new information about fundamental topics, ranging from the formation and evolution of the first massive galaxies and protoclusters to the detection of radio emission from Jupiter-like exoplanets \citep[see][for a summary of the science cases]{deGasperin2021}. Moreover, with the unprecedented sensitivity of LOFAR in this hitherto poorly studied spectral region, there is the possibility of serendipitous discoveries based on the large number of faint ultra-steep spectrum sources that will be detected in the LOFAR large-sky low-frequency surveys. 

\begin{figure}
\centering
 \includegraphics[width=0.5\textwidth]{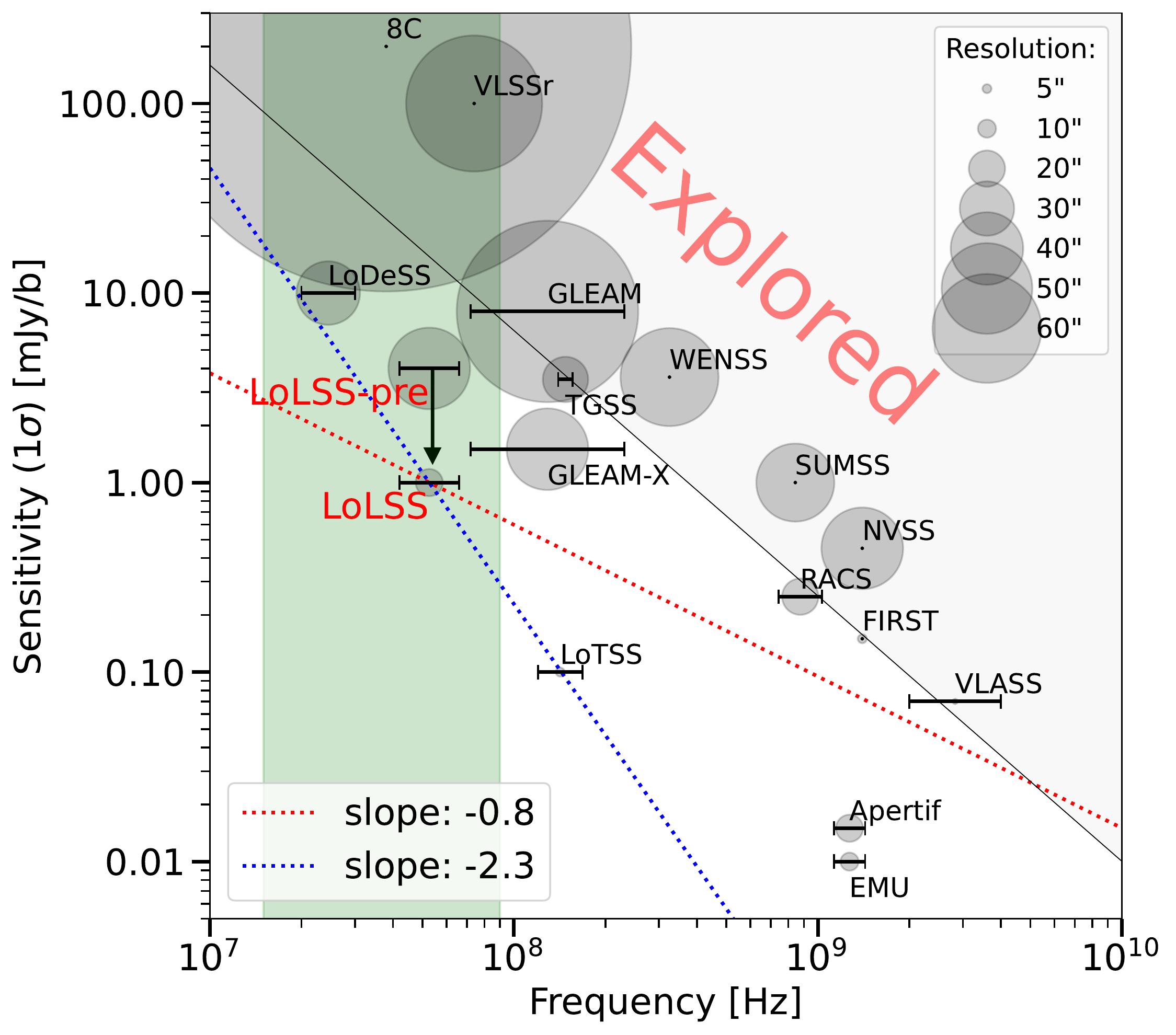}
 \caption{Comparison of sensitivity for a number of completed and ongoing wide-area radio surveys. The diameters of the grey circles are proportional to the survey resolution as shown in the top right corner. The green region shows the frequency range covered by the LOFAR LBA system. Data presented in this paper are labelled as `LoLSS', whilst the preliminary release of the \lol{} survey is labelled as `LoLSS-pre'. For sources with a very steep spectral index ($\alpha \lesssim -2.3$), \lol{} is the most sensitive survey currently available. References: 8C \citep{Rees1990}; GLEAM \& GLEAM-X \citep[GaLactic and Extragalactic All-sky Murchison Widefield Array survey;][]{Hurley-Walker2017, Hurley-Walker2022}; TGSS ADR1 \citep[TIFR GMRT Sky Survey - Alternative Data Release 1;][]{Intema2017}; VLSSr \citep[VLA Low-frequency Sky Survey redux;][]{Lane2014}; FIRST \citep[Faint Images of the Radio Sky at Twenty Centimetres;][]{Becker1995}; NVSS \citep[1.4 GHz NRAO VLA Sky Survey;][]{Condon1998}; WENSS \citep[The Westerbork Northern Sky Survey;][]{Rengelink1997}; SUMSS \citep[Sydney University Molonglo Sky Survey;][]{Bock1999a}; RACS \citep[Rapid ASKAP Continuum Survey;][]{McConnell2020}; Apertif (Adams et al. in prep.); EMU \citep[Evolutionary Map of the Universe;][]{Norris2011}; VLASS \citep[VLA Sky Survey;][]{Lacy2020}; LoTSS \citep[LOFAR Two-metre Sky Survey;][]{Shimwell2017}; LoDeSS (LOFAR Decameter Sky Survey; Groeneveld et al. in prep.).}
 \label{fig:survey}
\end{figure}

The high survey speed of modern radio telescopes, combined with increased complexity in dealing with extremely large data rates, shifted the observing strategy of several radio observatories in favour of large-scale surveys \citep[][see Fig.~\ref{fig:survey}]{Norris2017}\footnote{Examples of these are: the Evolutionary Map of the Universe \citep[EMU;][]{Norris2011,Norris2021}, the Polarization Sky Survey of the Universe’s Magnetism \citep[POSSUM;][]{Gaensler2010}, the APERture Tile In Focus surveys (APERTIF; Hessin in prep.), the GaLactic and Extragalactic All-Sky MWA-eXtended survey \citep[GLEAM-X][]{Hurley-Walker2017,Hurley-Walker2022}, and the Karl G. Jansky Very Large Array Sky Survey \citep[VLASS;][]{Lacy2020}.}. The LOFAR survey team set up a multi-tiered strategy aimed at covering a large portion of the northern sky at both HBA and LBA frequencies:
\begin{itemize}
 \item The LOFAR Two-metre Sky Survey \citep[LoTSS;][]{Shimwell2019,Shimwell2022} is a survey of the entire northern sky in the frequency range $120-168$ MHz. The survey reaches a sensitivity of about 80~\mujybeam{} at a resolution of $\approx 6\arcsec$. Several deeper fields with much longer exposures were also included in the LoTSS survey within the frame of the LoTSS deep field program \citep[see e.g.][]{Tasse2021, Williams2021, Sabater2021}. Although LoTSS does not use them, the survey data include international stations that allow imaging at 0.3\arcsec{} resolution \citep{Morabito2021}. A re-imaged version of the LoTSS survey data including these stations is planned and the first high-resolution pilot study has recently been released \citep{Sweijen2022}.
 \item The LOFAR LBA Sky Survey \citep[LoLSS;][]{deGasperin2021} was designed as the LBA counterpart of LoTSS in the frequency range $42-66$ MHz. It aims to cover the northern sky at Dec~$>24\deg$, with a sensitivity of $\approx 1$~\mjybeam{} and a resolution of $\approx 15$\arcsec{}. Several selected deep fields were also observed within the survey framework, to achieve increased sensitivity and reach lower frequencies \citep[see e.g.][]{deGasperin2020b, Williams2021}.
 \item The LOFAR Decameter Sky Survey (LoDeSS; Groeneveld et al. in prep.) is an experimental survey designed to cover the northern sky at the lowest frequency range of the LBA system ($15-30$ MHz).
\end{itemize}

% history
Due to the low operating frequency and wide field of view, the main challenge that needs to be overcome to produce reliable LOFAR images is the removal of ionospheric-induced systematic effects \citep{Intema2009, Mevius2016, deGasperin2018a}. Because they vary across the field of view, these are direction-dependent effects. Therefore, self-calibration strategies previously adopted for correcting small-field interferometric data cannot be used to simply calibrate LOFAR interferometric data sets. For this reason more sophisticated tools needed to be developed or improved to reduce LOFAR data, such as WSClean \citep{Offringa2014} and DDFacet \citep{Tasse2018} for imaging, DP3 \citep{vanDiepen2018} and KillMS \citep{Tasse2014} for calibration, and LoSoTo for solution manipulation \citep{deGasperin2019}. The availability of these new tools required the adoption of new and specialised analysis strategies. The high sensitivity of HBA observations allowed for the first development of a working strategy based on facet calibration and imaging \citep{Weeren2016}, although LoTSS ended up using a faster approach based on the simultaneous derivation of solutions using KillMS and DDFacet \citep{Tasse2021}. In the LBA low signal-to-noise ratio regime, the facet calibration approach delivers stable results, so variations of this serial-solving approach were implemented in producing LoLSS \citep[e.g.][; see also Sec.~\ref{sec:reduction}]{Edler2022}.

% paper
The paper is organised as follows: in Sec.~\ref{sec:lolss} we describe the main observational parameters of the survey and its current status, updating what was initially presented in Sec. 3 of \cite{deGasperin2021}. In Sec.~\ref{sec:reduction} we give an update on the data reduction procedure initially presented in \cite{deGasperin2019, deGasperin2020a, deGasperin2021}. In Sec.~\ref{sec:release} we describe the LOFAR LBA  Sky Survey First Data Release and in Sec.~\ref{sec:results} we give quantitative estimation of the survey properties. A summary of the paper is in Sec.~\ref{sec:summary}. The spectral index $\alpha$ is defined as $S_\nu \propto \nu^{\alpha}$, with $S_\nu$ the flux density.

\section{The LOFAR LBA Sky Survey}
\label{sec:lolss}

\begin{table}
\centering
\begin{threeparttable}
\begin{tabular}{lc}
\hline\hline
Number of pointings & 1\,889 \\
Separation of pointings & 2.58\deg \\
Integration time (per pointing) & 3 h (8 h for HETDEX) \\
Frequency range & 42 -- 66 MHz \\
Array configuration & LBA OUTER / LBA SPARSE \\
Angular resolution & $\sim15$\arcsec \\
Noise level & $\sim 1-2$~\mjybeam \\
Time resolution & 1 s \\
-- after averaging & 4 s \\
Frequency resolution & 3.052 kHz \\
-- after averaging & 24.414 or 48.828 kHz \\
\hline
\end{tabular}
\end{threeparttable}
\caption{\lol{} observational setup.}\label{tab:specs}
\end{table}

% pointing scheme
A summary of the observational setup used for the LOFAR LBA Sky Survey is given in Table~\ref{tab:specs}. To cover the entire sky north of  declination $>24\deg$, the survey was divided into 1889 pointings that are centered on the same coordinates as those of the corresponding LoTSS pointings in order to simplify future combined analysis. Each observation used the LOFAR LBA multi-beam capability to observe simultaneously one calibrator chosen among 3C\,196, 3C\,295, and 3C\,380 and three target fields. All observations were carried out in exposure blocks of 1~h, so that each field was observed multiple times on different days to reduce the probability that a field could not be used due to bad ionospheric conditions. The various observations of the same field were scheduled to maximise the Local Sidereal Time coverage when the elevation of the target was above 60\deg. We set such a limit on the elevation in order to minimise both the suppression of the signal due to the dipole beam and the path length through the ionospheric layer that is pierced by the incoming radio waves.

%LST:
%40<dec<50: obs 1h - skip 1h - max elev - skip 1h - obs 1h
%50<dec<90: obs 2h - skip 1h - max elev - skip 1h - obs 2h

% frequency coverage
Multi-beam capability comes at the cost of a reduced bandwidth per beam. The total bandwidth that LOFAR can process is 96 MHz. In our case, each beam has allocated 24 MHz, i.e. 122 sub-bands (SB) of 0.195 MHz each, in order to cover the frequency range $42-66$~MHz. This is the most sensitive frequency range of the LBA system once the sky temperature and the dipole bandpass are taken into consideration \citep{vanHaarlem2013}. At the station level, a 30-MHz high-pass filter is applied to the signal path to suppress radio frequency interference (RFI) reflected by low ionospheric layers at $<20$~MHz.

% modes
There are three operating modes for LOFAR LBA:
\begin{itemize}
 \item LBA\_INNER, for which the innermost 48 dipoles of each station are used. This mode gives the largest beam size at the cost of a reduced sensitivity. The calibration of the individual station dipoles in this mode is less effective than for the other modes due to mutual coupling and increased response to Galactic emission during the station calibration procedure. The effective size of the station in this mode is 32 m, which corresponds to a primary beam full width at half maximum (FWHM) of 10\deg{} at 54 MHz.
 \item LBA\_OUTER, where only the outermost 48 dipoles of each station are used. This mode minimises the coupling between dipoles and reduces the beam size. The effective size of the station in this mode is 84 m with a primary beam FWHM of 3.8\deg at 54 MHz.
 \item LBA\_SPARSE (ODD or EVEN), where half of the dipoles, distributed across the station, are used. This mode results in an intermediate performance between LBA\_INNER and LBA\_OUTER, with a suppression of the magnitude of the side-lobes compared to the latter. The effective size of the station in this mode is about 65 m, which results in a primary beam FWHM of 5.5\deg{} at 54 MHz.
\end{itemize}

Planned updates of LOFAR will triple the processing power at the stations enabling the use of all LBA dipoles simultaneously (LBA\_ALL mode), effectively doubling the station collecting area with a FWHM similar to LBA\_SPARSE (Hessels et al. in prep.). Until the end of 2020, LoLSS observations were carried out using the LBA\_OUTER mode. This includes all observations of the first release discussed in this paper. After commissioning LBA\_SPARSE, we switched to that mode to increase the FoV and therefore the survey speed while reducing the problem of unmodelled flux spilling into the side-lobes of the primary beam.

% pointing scheme
The pointing schemes used for LoLSS followed a spiral pattern starting from the north celestial pole, with pointing positions determined using the \cite{Saff1997} algorithm and separated by 2.58\deg. With such a pointing scheme, \lol{} required 1889 pointings. Assuming circular beams and LBA\_OUTER, this separation provides a distance between pointing centres of FWHM~$/\sqrt{2}$ at the highest survey frequencies and better than FWHM~$/2$ at the lowest. With LBA\_SPARSE, the overlap of the fields is even greater.

% stations
At present, the International LOFAR Telescope has 24 core stations (CS), 14 remote stations (RS), 14 international stations (IS) and two more under construction. The CSs are spread across a region of radius $\sim2$ km and provide 276 short baselines. The RS are located within 70 km from the core and their longest baseline provides a resolution of $\sim15\arcsec{}$ at 54 MHz. LoLSS makes use of CS and RS, whilst IS data were not recorded to keep the size of the data set manageable.

% sensitivity
The aim of LoLSS is to reach a sensitivity of $\sim 1$~\mjybeam. With the LBA system, this requires $\approx 8$~hrs of integration time at the optimum declination. However, because of the strong overlap of LBA\_SPARSE fields, this integration time can be relaxed to about 3 hours per field. In most cases, the final noise is limited by ionospheric conditions and dynamic range while confusion limit is expected to be $< 200$~\mujybeam{} \citep{Condon2012}. Experiments with comparable total integration times indicate that the rms noise ranges between 1 and 2~\mjybeam{} \citep[e.g.][]{deGasperin2020a}. In the preliminary release, in which the direction-dependent errors were not corrected, the noise ranged between 4 and 5~\mjybeam{} \citep{deGasperin2021}, compared with a median noise of 1.6~\mjybeam{} for this release (see Sec.~\ref{sec:sensitivity}).

% averaging
The time and frequency resolution were chosen to balance the data size and the effect of time/frequency smearing at the edges of the field of view. A time resolution of a few seconds is also necessary to track fast evolving ionospheric variations. Data were initially recorded at 1 s and 3.052 kHz resolution to properly identify fast and narrow band RFI \citep{Offringa2010}. The high resolution is also needed to remove bright sources (Cygnus A and Cassiopeia A) from the far side lobes \citep{deGasperin2019}. Data were then averaged to 2 s and 48.828 kHz for LBA\_OUTER observations and to 4 s and 24.414 kHz for LBA\_SPARSE. The decrease in frequency averaging was applied to keep the frequency smearing to less than 5\% at the edge of the Field of View (FoV). Time smearing is less of a problem ($<1\%$) and an averaging time of 4 s was selected to keep the data volume low.

\begin{figure}
\centering
 \includegraphics[width=.5\textwidth]{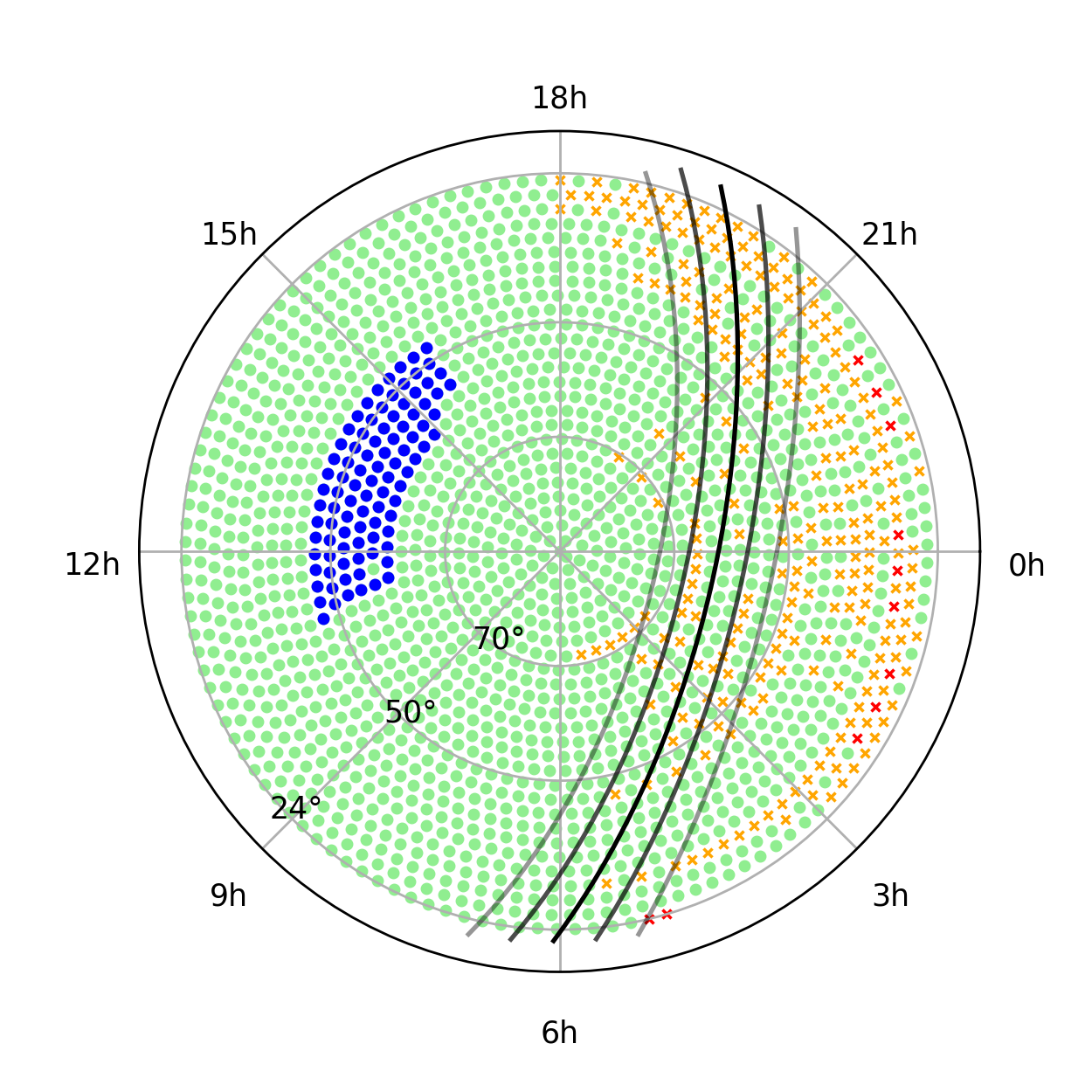}
 \caption{Current and planned sky coverage of \lol{}. Each marker represents a pointing of the full survey. The pointings of the region presented in this paper are the blue circles. Green circles and yellow crosses represent pointings with archived observations (3 hrs for green, 1 or 2 hrs for yellow). Red crosses will be observed after 2022. Solid lines show the position of the Galactic plane with Galactic latitude: $-10 \deg, -5\deg, 0\deg, 5\deg, +10$\deg.}
 \label{fig:coverage}
\end{figure}

\subsection{Survey status}
\label{sec:survey_status}

The initial observations of LoLSS, those covering the region of the Hobby--Eberly Telescope Dark Energy Experiment \citep[HETDEX][]{Hill2008} spring field, were obtained in 2017 and 2019. After proving the feasibility of the survey, the bulk of the observations started in 2020 and are expected to be completed by mid 2023. In Fig.~\ref{fig:coverage}, we give an overview of the observing status as of October 2022. At the time of writing 1507 (79.8\%) pointings have been observed with at least 3 hrs of data, 244 (12.9\%) have been observed for 2 hrs and 112 (5.9\%) for 1 hrs, while 26 (1.4\%) pointings still have no observations. Pointings that were unusable due to hardware issues or that were taken during particularly bad ionospheric conditions (about 15\%) were identified. They are not included as valid data in the above summary and re-observations are on-going.

\section{Data reduction}
\label{sec:reduction}

The images presented in the first release of LoLSS were obtained through a full re-run of the calibration process, neglecting the partially calibrated data of the preliminary release. The data reduction was carried out at the Hamburg Observatory using five computing nodes with 32 cores each, 256 GB of memory and 20 TB of storage in local drives. A centralised database kept track of the various jobs and coordinated the nodes. The computation was carried out in an ad-hoc environment based on Ubuntu 20.4 for which a Docker container is available\footnote{See \url{https://github.com/revoltek/LiLF}.}.

For the data reduction, an automated Pipeline for LOFAR LBA (PiLL)  that is described in \cite{deGasperin2019} for the calibrator steps, and \cite{deGasperin2020a} for the target steps, was employed. The direction-dependent portion was rewritten compared with that used to produce the first publication of a thermal-noise limited LOFAR LBA image presented in \cite{deGasperin2020a}. The main differences and a brief overview of the full data reduction procedure are described below.

\subsection{Pre-processing pipeline}

Immediately following the observations, the following steps were carried out by the LOFAR Observatory at ASTRON: 1. Flagging of RFI with AOflagger \citep{Offringa2012}, 2. Subtraction (``demixing'') of the ultra-bright sources Cygnus A and Cassiopeia A using the algorithm described in \cite{VanderTol2007} and 3. Averaging of the data to the relevant time and frequency resolution. As described above, these were  4 s and 24.414 or 48.828 kHz per channel depending on the observing mode, with the former used with LBA\_SPARSE and the latter with LBA\_OUTER. The data were then ingested into the LOFAR Long Term Archive \citep{Renting2012}.

\subsection{Calibrator pipeline}

The calibrator pipeline was then applied to data of the beam pointing at the calibrator source. The calibrator was chosen from the bright compact sources 3C\,196, 3C\,295, and 3C\,380, depending on which was at higher elevation during the observation. Since our observations did not include data from international stations, the resolution is limited to 15\arcsec{}.  We therefore could use simple models for the calibrators. 3C\,196 was parameterised by four point sources and 3C\,295 by two point sources. However, 3C\,380 is well-resolved even at 15\arcsec{} and a more detailed model for this source was obtained after self-calibrating it using LOFAR LBA data including international stations \citep{Groeneveld2022}. The total flux density of all models were rescaled to match the \cite{Scaife2012} flux density scale.

The calibrator pipeline then isolated the polarisation misalignment introduced by the station calibration table. This is an artificial delay between the two polarisation components that is constant in time and varies with station. The Faraday rotation was removed by converting the data set to circular polarisation and measuring the separation between right and left circularly polarised polarisation, which has a frequency dependency $\propto 1/\nu^2$. This approach has the advantage that it does not depend on the sky model. Faraday rotation was derived solely by measuring the relative misalignment between the two polarisations (Groeneveld et al. in prep). Finally, a time-independent bandpass was derived together with a time and frequency dependent scalar phase that encodes differential station delays due to misaligned clocks and differential ionospheric delays. The two effects are theoretically separable using their different frequency dependency \citep{Mevius2016}. However, in the LOFAR LBA low signal-to-noise ratio regime, strong cross-contamination was found to frequently persist. 

\subsection{Target pipeline: direction-independent}

Data from each target field were corrected using solutions derived from the simultaneous calibrator observation. The solutions applied were: polarisation alignment, bandpass and scalar phase calibration. Here the polarisation alignment, bandpass and the clock part of the phases are instrumental systematic effects that are direction-independent, and are therefore fully corrected for using the calibrator solutions. However, the ionospheric part of the scalar phases corrupts the data with the effect of the ionosphere in the calibrator direction. Finally, the theoretical element beam of LOFAR LBA dipoles was applied and the data were combined into a single measurement set.

At this stage the direction-independent self-calibration can begin. The data were initially calibrated using a sky model derived from the combination of data from TGSS \citep{Intema2017}, NVSS \citep{Condon1998}, WENSS \citep{Rengelink1997}, and VLSS \citep{Lane2014}. This enabled the spectral index of each source to be estimated and its flux density to be extrapolated to the LBA frequency range. Sources with flux densities smaller than 1 Jy at 60 MHz were discarded from the model to speed up the prediction of the visibilities.

During self-calibration, corrections for three effects were derived. The first is a direction-independent Faraday rotation, derived using the same procedure described for the calibrator field. Secondly, the direction-independent total electron content (TEC) of the ionosphere was estimated for each station. This was done using slow-varying (1 minute) solutions for the stations within 10 km from the central core of LOFAR (the `Superterp') and fast-varying (4 seconds) solutions for the other stations. Thirdly, station-independent amplitude solutions were obtained on all core stations in an attempt to derive the second-order corrections on the analytical beam model. These solutions were then applied to data from the core and remote stations.

During the process, a wide-field, low-resolution image was produced. This image was used to identify strong sources present outside the first null. These are particularly problematic in the LBA\_OUTER mode for which the first side-lobe is prominent. Sources found outside the main lobe were then subtracted from the visibilities. An image covering the entire field of view out to the first null was finally produced. The images produced at this stage of the process are similar to those presented in the preliminary release \citep{deGasperin2021}.

\begin{figure}[htb!]
 \centering
 \includegraphics[width=.45\textwidth]{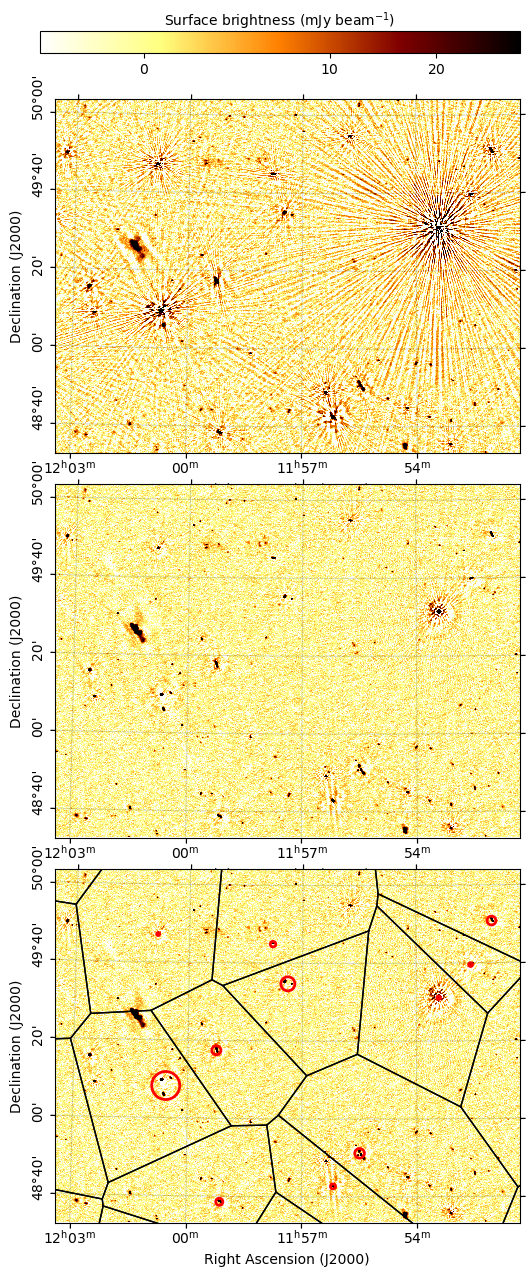}
 \caption{A region of the LBA survey with only direction-independent error corrections applied (top) and with both direction-independent and direction-dependent error corrections applied (middle and bottom). In the bottom panel we identify the direction-dependent calibrators used (sources within red circles) and the relative facets. Inclusion of direction-dependent error corrections results in a sensitivity improvement by a factor $\sim2$.}
 \label{fig:diedde}
\end{figure}

\subsection{Target pipeline: direction-dependent}
\label{sec:pipelinedd}

At this point, differential (w.r.t. the average TEC) direction-dependent ionospheric errors caused by local TEC variations across the field and second-order direction-dependent beam errors need to be addressed. The pipeline identifies suitable direction-dependent calibrators using the image produced at the end of the previous step. These calibrators are grouped using machine learning techniques based on mean shift clustering \citep{Cheng1995}, and finally selected based on: 1. a minimum apparent flux density, with $S_\nu > 1.0$~Jy at $\nu = 60$~MHz assuming a spectral index $\alpha=-0.8$ and 2. avoiding clusters of sources for which at least one source, that contributes with more than 20\% to the cluster flux density, is extended. Here a source is considered extended if it has an integrated to peak flux density ratio $S_{\rm I} / S_{\rm P} > 4$. Extended sources are discarded as their signal does not contribute substantially to the longest baselines, that are critical to calibrate the most remote stations. See Fig.~\ref{fig:diedde} for examples of direction-dependent calibrators.

The data set was further averaged in time to resolution of 8 s to decrease the data volume. For each calibrator, in decreasing order of flux density, all other sources were subtracted from the data set. The data set is phase-shifted in the calibrator direction and averaged to channels of $2 \times 192$~kHz and to 16 or 32 seconds if the calibrator flux density is lower than 10 or 4 Jy, respectively. This averaging is possible as most of the ionospheric effects are already removed at this stage and frequency/time smearing is negligible in the small region surrounding the calibrators. The differential beam effect is applied to the averaged data set and solutions are derived in several cycles of self-calibration on the calibrator sources. In each cycle, fast scalar phases are derived for remote stations, with solutions forced to be smooth in frequency through the solver. The loop continues as long as the image noise and dynamic range improve. If the process arrives to the fourth cycle, for sources with an apparent flux density $>5$ Jy at 54 MHz, amplitude solutions are derived in two steps. Firstly, slow amplitude solutions for six frequency chunks are derived forcing all stations to have the same solution. These amplitude variations reflect imperfect element beam modelling that are expected to be the same for each station. Secondly, even slower solutions are derived for each antenna forcing the solver to keep the solutions smooth in frequency. All amplitude solutions are normalised so as not to affect the overall flux scale. During the self-cal cycles the calibration steps are carried out using DP3 \citep{vanDiepen2018}, the imaging using WSClean \citep{Offringa2014,Offringa2017}, and the handling of the solutions using LoSoTo \citep{deGasperin2019}.

Solutions from direction-dependent calibrators that resulted in a lowering of the local noise and increase in the dynamic range are preserved and combined in an imaging call using DDFacet \citep{Tasse2018}. The imager applies the solutions to each facet without attempting a smooth transition between them. Together with the solutions, a time-dependent model of the beam is also applied to each facet. A mask is also generated from the best available image and passed to DDFacet to include all extended emission in the deconvolution process, providing a good fidelity. A circular synthesized beam of 15\arcsec{} is used as restoring beam. An example of the final result of the imaging step is visible in Fig.~\ref{fig:diedde}. Finally, the entire direction-dependent calibration process is repeated but using the improved model and accounting for the direction-dependent solutions during the subtraction steps. At the end of the second main cycle, DDFacet produces I- and V-Stokes images as well as a 6-channel cubes which will be used to derive in-band information.

\section{Public data release}
\label{sec:release}

Here we describe the data products released to the community. They consist of images and source catalogues. The present products cover a region around the HETDEX spring field (RA: 11 h to 16 h and Dec: 45\deg{} to 62\deg{}; area: 650 deg$^2$), which is 5\% of the sky area to be covered by the LoLSS survey.

\subsection{Mosaic images}
\label{sec:mosaic_image}

\begin{figure*}
 \centering
 \includegraphics[angle=90,height=24cm]{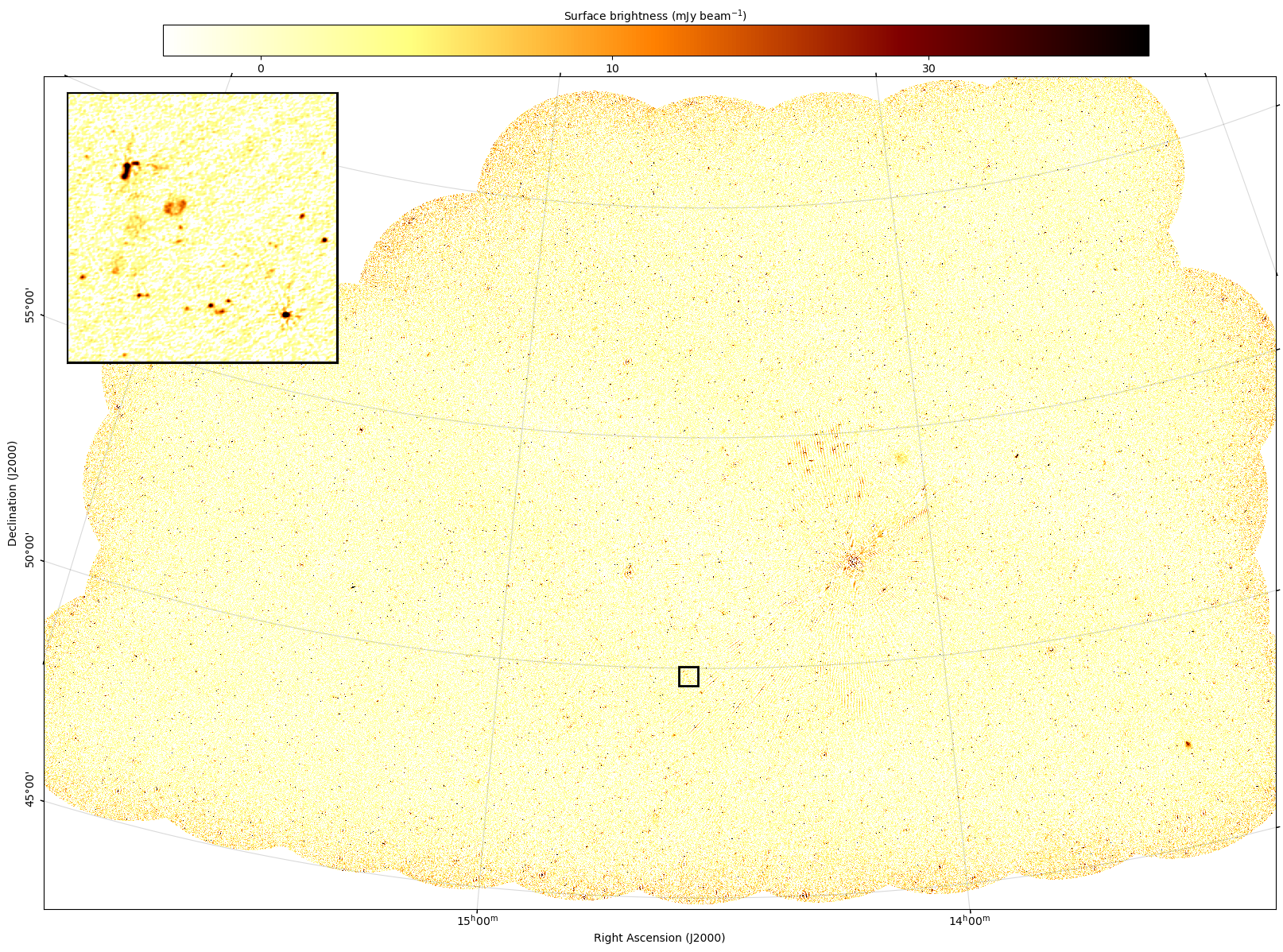}
 \caption{Mosaic image of the total intensity emission in the eastern half of survey area (resolution: 15\arcsec). The small panel is a zoom of the marked region.}
 \label{fig:mosaic}
\end{figure*}

\begin{figure*}
 \centering
 \includegraphics[angle=90,height=24cm]{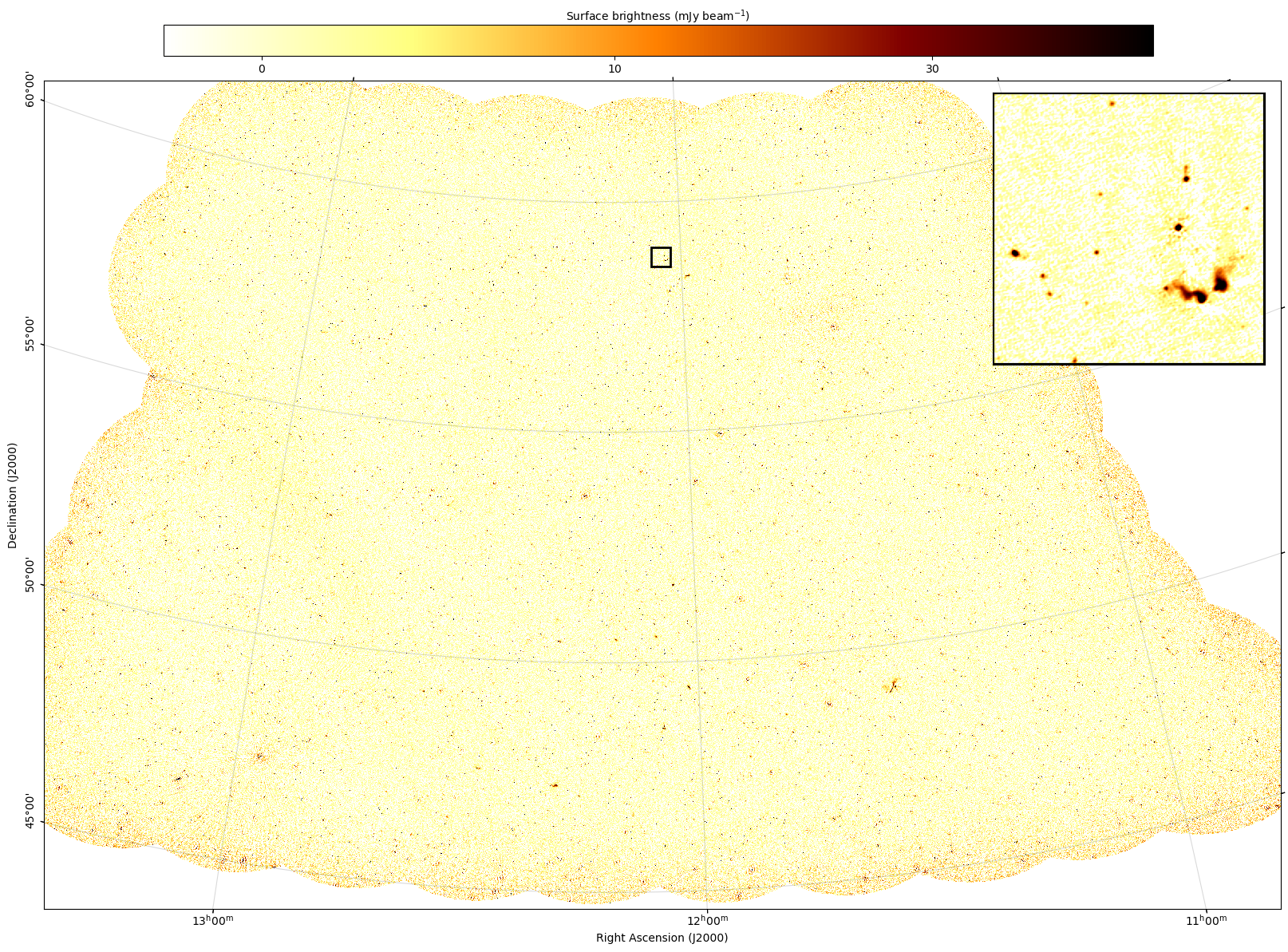}
 \caption{Mosaic image of the total intensity emission in the western half of survey area (resolution: 15\arcsec). The small panel is a zoom of the marked region.}
 \label{fig:mosaic2}
\end{figure*}

\begin{figure*}
\centering
 \includegraphics[width=\textwidth]{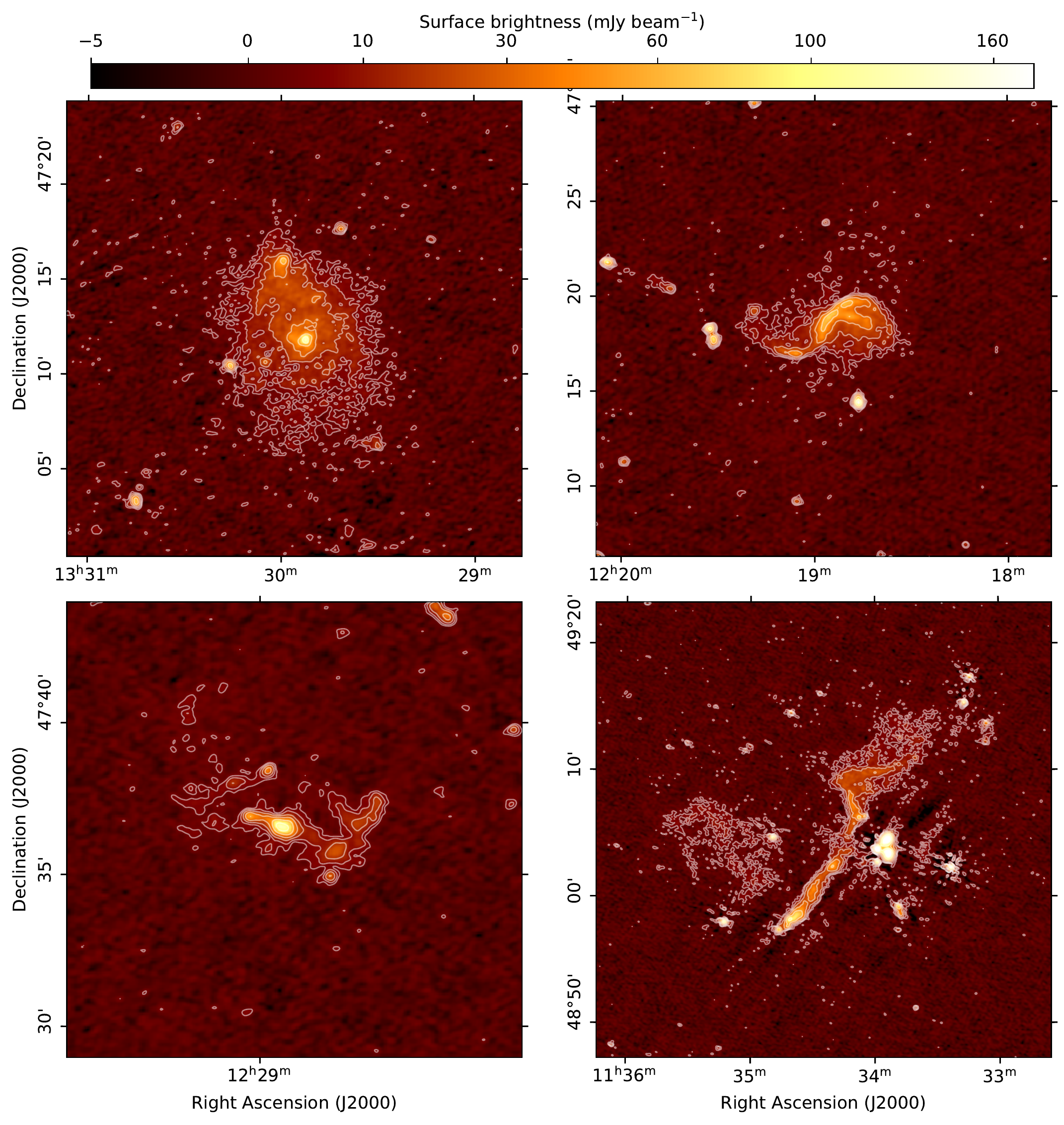}
 \caption{Some examples of extended sources in the data release published with this paper. From top-left to bottom-right M\,51, M\,106, the galaxy cluster Abell~1550, and the complex blend of emission coming from both active galactic nuclei (AGN) and diffuse sources in the intra-cluster medium of Abell~1314. Contours are at $(4.5,8,15,30,50,100)$~\mjybeam{}, so that the first contour is approximately at $3\sigma$.}
 \label{fig:diffuse}
\end{figure*}

We produced one mosaic of $3.3\deg\times3.3\deg$ around each of the 95 pointing centres of the HETDEX region. Mosaics were made by taking all neighbouring pointing images with centre distances $< 6\deg$, and reprojecting them into the frame of the central pointing. The images were then corrected for any global astrometric shift derived by cross-matching isolated and compact sources against an external catalog. For this purpose, we employed the FIRST survey. The distribution of the corrections applied has a standard deviation of 2\arcsec{} for both RA and Dec. Finally, the reprojected images' data were aggregated to create a single output image using inverse variance weighting, with the weights of each pixel of each input image calculated dividing the beam attenuation in that pixel by the average rms noise of the input image.

%in-band, Vstokes
The same process was repeated for each channel of the final image cube, producing mosaics at 44, 48, 52, 56, 60, and 64 MHz for which only the source catalogues will be provided. For the combined image at 54 MHz we also produced and released Stokes\,V mosaics. Together with the 95 mosaics at 54 MHz, we also prepared two large mosaics covering the east and west part of the HETDEX region: these are shown in Fig.~\ref{fig:mosaic} and \ref{fig:mosaic2}. Within each figure, a small panel shows a blowup of a region of the mosaic to illustrate the data quality. A few examples of extended sources, showing the fidelity of the final product, are presented in Fig.~\ref{fig:diffuse}. The rms noise of the final images is on average 1.55~\mjybeam{}, the sensitivity of the survey is discussed in detail in Sec.~\ref{sec:sensitivity}. 

\begin{figure*}[htb]
\centering
 \includegraphics[width=\textwidth]{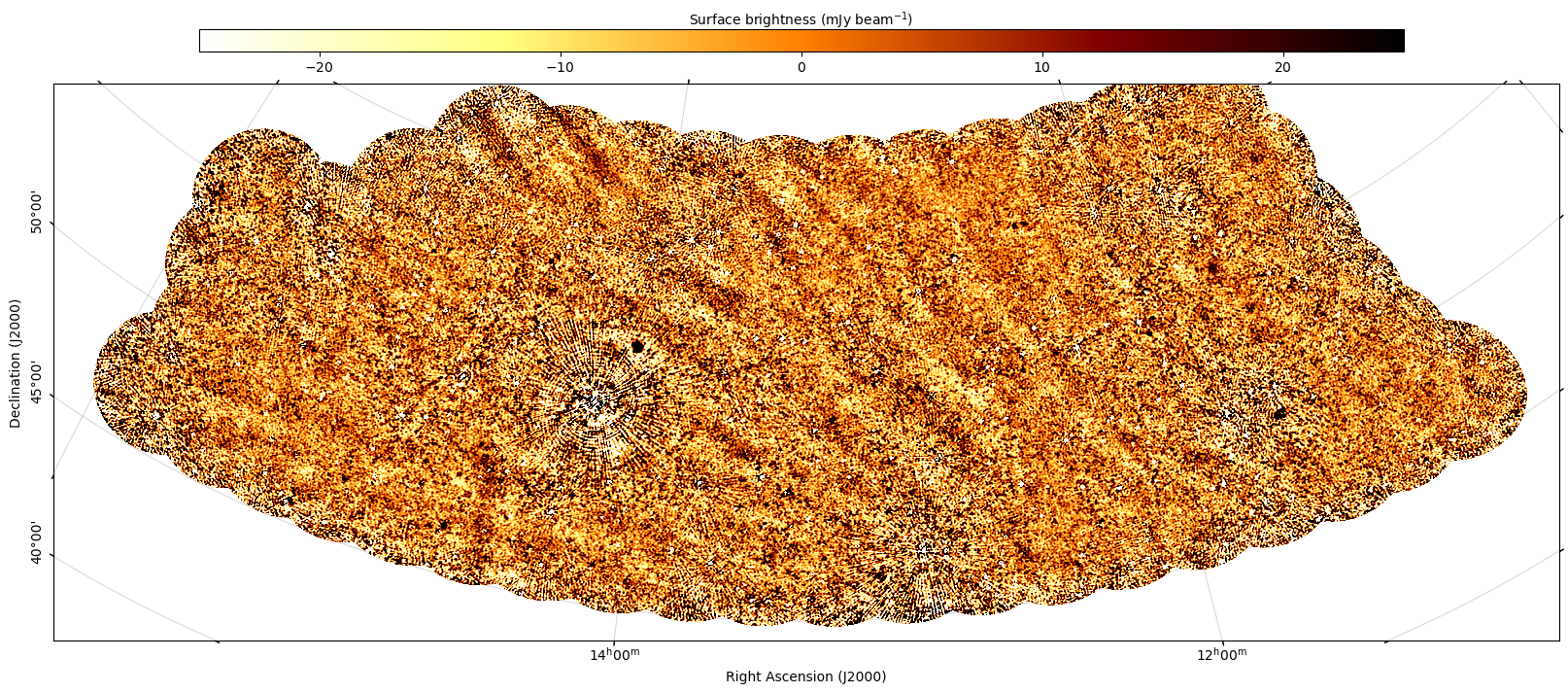}
 \caption{Low-resolution (3\arcmin), source-subtracted map of the survey area. Large and locally aligned stripes of emission are visible crossing the region. The same structures were identified in the corresponding LoTSS images, which were calibrated independently with a different calibration code, demonstrating that these features are not artefacts.}
 \label{fig:mosaic-lr}
\end{figure*}

% low-res
Finally, a low-resolution image was produced starting from direction-independent error corrected data. For each field, the detected sources were modelled taking into account direction-dependent effects and subtracted from the visibilities. Then the data sets were imaged by Gaussian tapering the weights to reach a resolution of 180\arcsec{}. In this case, the direction-dependent error correction is less relevant as only short baselines were preserved (max 5 km). The mosaic image is presented in Fig.~\ref{fig:mosaic-lr}. The image shows the presence of large angular-scale structures that were independently detected also in LoTSS data and are likely of Galactic origin (Oei et al. in prep.). We caution the reader that the surface brightness of such structures might be strongly biased due to their large angular size combined with missing short baselines. Because of the presence of the extended emission in all regions of the map, it is difficult to estimate the background rms noise. Measuring it in a few regions where the emission appears less dominant, we find an rms noise of about $10-15$~\mjybeam. Images described in this section are available at \url{www.lofar-surveys.org/lolss.html}.

\subsection{Source catalogues}
\label{sec:catalogue}

% catalogue production
Source catalogues were generated with PyBDSF \citep{Mohan2015}. The source extraction procedure used a $4\sigma$ detection threshold to find islands of emission with a $5\sigma$ threshold on the brightest pixel. The significance of the emission in each pixel is measured against the local rms noise. To reduce the number of false positive detections, we use an adaptive rms box size that shrinks around sources with pixels above 50 times the local rms. We saved both the Gaussian components used to fit the source shape and the source catalogue. Rms images were also saved to be used in subsequent analysis. Each run of the source finder was performed on a $3.3\deg\times3.3\deg$ mosaic; this produced 95 Gaussian components and 95 source catalogues with significant overlap. The combined source catalogue was then produced by merging the 95 separate source catalogues and for each retaining only the sources closest to the centre of that particular mosaic image. The combined Gaussian components catalogue retained only Gaussian components associated with sources in the combined source catalogue. The combined source catalogue contains 42\,463 entries and the combined Gaussian components catalogue contains 53\,377 components. An overview of the catalogues' columns is given in Table~\ref{tab:cat}. The catalogues are available at \url{www.lofar-surveys.org/lolss.html}.

\begin{table}
\centering
\begin{threeparttable}
\begin{tabular}{lc}
\hline\hline
Column name & Format/Unit \\
\hline
Source\_name & e.g. LOL1J104657.0+482723 \\
Gaus\_id & Integer \\
E\_RA & arcsec \\
DEC & deg \\
E\_DEC & arcsec \\
Total\_flux & mJy \\
E\_Total\_flux & mJy \\
Peak\_flux & mJy beam$^{-1}$ \\
E\_Peak\_flux & mJy beam$^{-1}$ \\
Maj & arcsec \\
E\_Maj & arcsec \\ 
Min & arcsec \\
E\_Min & arcsec \\ 
PA & deg \\
E\_PA & deg \\
DC\_Maj & arcsec \\ 
E\_DC\_Maj & arcsec \\ 
DC\_Min &  arcsec \\
E\_DC\_Min & arcsec \\
DC\_PA & deg \\
E\_DC\_PA & deg \\
Isl\_rms & mJy beam$^{-1}$ \\
S\_Code & S, M or C \\
Mosaic\_id & e.g. p164+47 \\
\end{tabular}
\end{threeparttable}
\caption{Columns and units present in the catalogue. The columns are: source identifier (Source\_name), Gaussian identifier (present only in the Gaussian components catalogues as derived by the source finder. Although a source can have multiple Gaussian components, the combination of Mosaic\_id and Gaus\_id is unique), J2000 right ascension (RA), J2000 declination (Dec), integrated flux density (Total\_flux), peak brightness (Peak\_flux), major axis (Maj), minor axis (Min), position angle (PA), deconvolved major axis (DC\_Maj), deconvolved minor axis (DC\_Min), deconvolved position angle (DC\_PA), local noise at the source location (Isl\_rms), type of source as classified by PyBDSF (S\_Code -- here 'S' indicates an isolated source fit with a single Gaussian; 'C' represents sources fit by a single Gaussian but within an island of emission that contains other sources; and 'M' is used for sources that are extended and fitted with multiple Gaussians), the mosaic identifier (Mosaic\_id). The errors in the catalogue are the uncertainties obtained from the PyBDSF source fitting. Additional uncertainties on the source extensions, astrometry and flux density scale are discussed in the text.}\label{tab:cat}
\end{table}

% single chan
The same process described above was repeated for each channel image derived from the final image cubes, therefore providing 6 additional source and Gaussian components catalogues. They have the same structure as the combined catalogue but because they are derived from images made with a sixth of the bandwidth, they contain fewer sources (see Table~\ref{tab:catentries}).

\begin{table}
\centering
\begin{threeparttable}
\begin{tabular}{lccc}
\hline\hline
Catalogue & Central   & Number    & Number of \\
          & frequency & of source & gaussian components \\
\hline
Combined  & 54 MHz & 42\,463 & 53\,377 \\
Chan 00   & 44 MHz & 22\,689 & 27\,067 \\
Chan 01   & 48 MHz & 24\,511 & 29\,213 \\
Chan 02   & 52 MHz & 25\,488 & 30\,443 \\
Chan 03   & 56 MHz & 25\,456 & 30\,363 \\
Chan 04   & 60 MHz & 23\,894 & 28\,380 \\
Chan 05   & 64 MHz & 21\,659 & 25\,636 \\
\end{tabular}
\end{threeparttable}
\caption{Catalogue frequencies and numbers of entries.}\label{tab:catentries}
\end{table}

\subsubsection{Completeness and false positives}

% completeness
To evaluate the catalogue completeness, we adopted the procedure outlined by \cite{Heald2015}. For this purpose, we used the residual mosaic images created by PyBDSF after subtracting the Gaussian components derived during the source detection. These images include information about the distribution of the rms noise and artifacts, and can therefore be used to inject artificial sources into the mosaics in order to assess to what level they can be retrieved by the source finder. To study this, we injected a population of 10\,000 point sources into each of the residual mosaic images, randomly distributed, with flux densities ranging between 3 mJy (two times the rms noise) and 10 Jy, and following a number count power-law distribution of $dN/dS \propto S^{-1.6}$ \citep{Wilman2008}. We then attempted to detect these sources using PyBDSF with the same parameters used in compiling the real catalogue. The process was then repeated 20 times to decrease sample noise.

\begin{figure}
\centering
 \includegraphics[width=.5\textwidth]{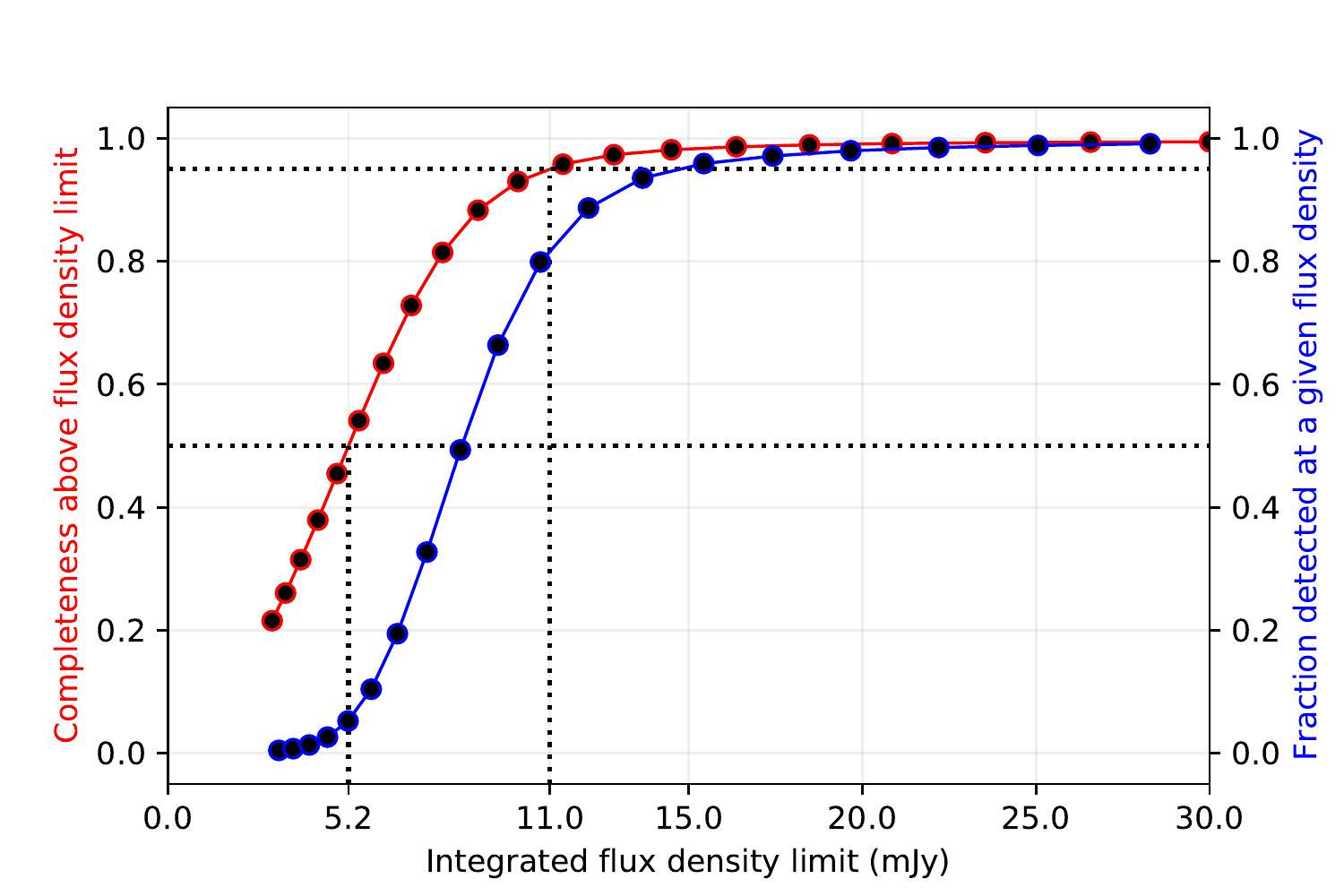}
 \caption{In red the completeness function above a certain flux density limit and in blue the fraction of detected sources at that flux density. The dotted lines show the 50\% and the 95\% completeness at 5.2 and 11 mJy respectively.}
 \label{fig:completeness}
\end{figure}

A source was accepted as detected if it was found to be within 45\arcsec{} (3 times the synthesised FWHM beam) of its input position and whose difference between the measured and simulated flux densities was within ten times the error on the recovered flux density. In Fig.~\ref{fig:completeness} we show the results for all 95 mosaics combined. We found a 50 percent probability of detecting sources at 8.5 mJy and 95 percent probability of detecting sources at 15 mJy. This is about three times better than for the preliminary release \citep{deGasperin2021}. The same figure also reports the fraction of recovered sources above a certain flux density. This shows that our catalogue is 50 percent complete over 5.2 mJy and 95 percent complete over 11 mJy, although we note that the cumulative completeness values depend on the assumed slope of the input source counts.

\begin{figure}
\centering
 \includegraphics[width=.5\textwidth]{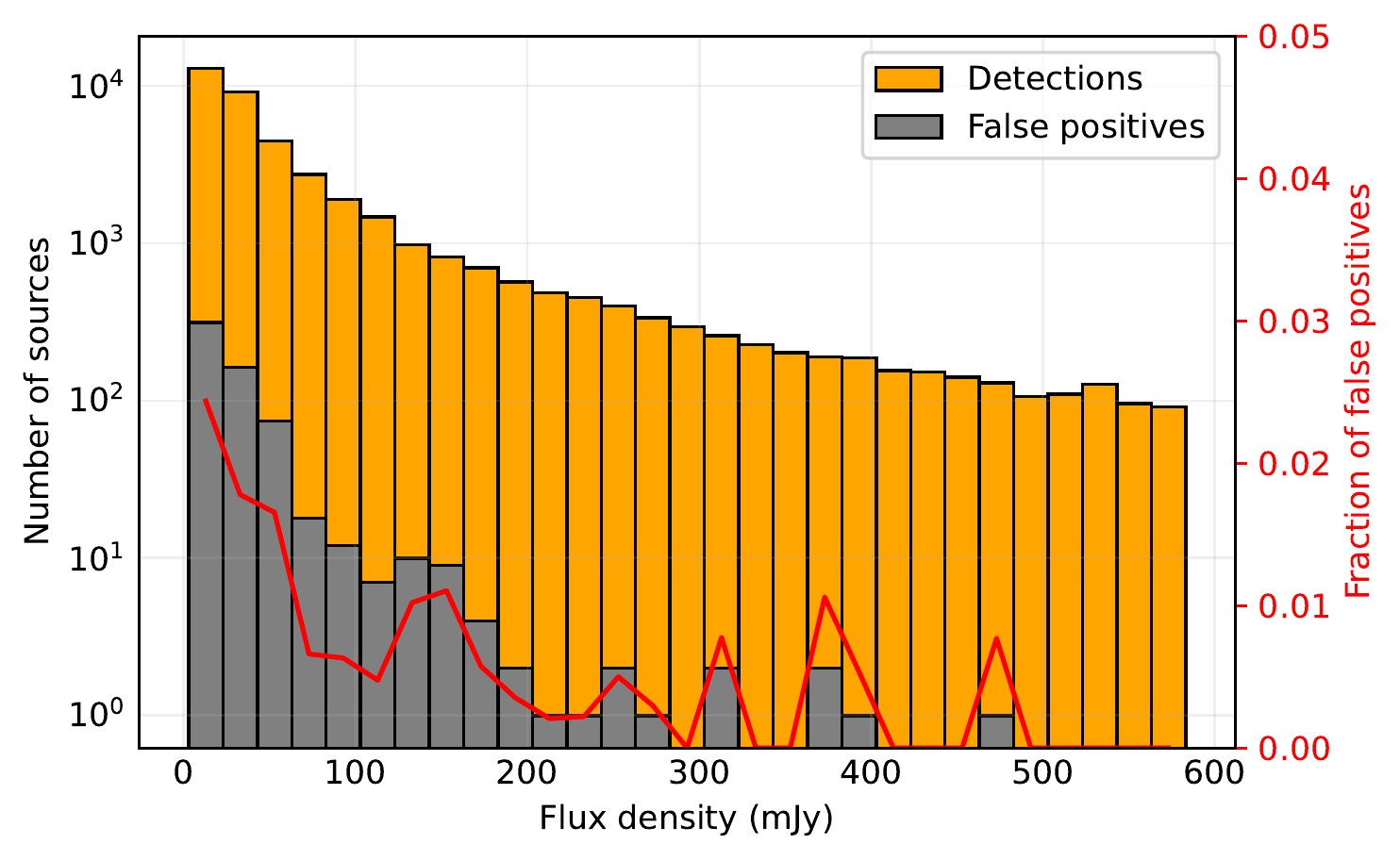}
 \caption{Distribution of detected sources in the LoLSS catalogue (yellow) and distribution of false positive (gray). In red the fraction of false positives per flux density bin.}
 \label{fig:falsepositive}
\end{figure}

% false positives
To assess the number of false positive detections, we inverted the pixel values of the residual mosaic images. This resulted in negative pixels due to noise and artefacts becoming positive. PyBDSF was then run with the nominal parameters used to generate the source catalogue. However, since real sources were not present in the images, the rms map would have been different from the original run, therefore the code was forced to use the rms map generated from the original images. On this basis, we estimated that 1.47\% of the sources in the LoLSS catalogue released here are due to artifacts. A distribution of the expected false positives as a function of flux density is shown if Fig.~\ref{fig:falsepositive}. Most of the false positives are below 300 mJy. Their distribution reaches relatively high flux densities and shows that they are principally caused by artefacts surrounding bright sources.

\section{Results}
\label{sec:results}

In this section we present the properties of the first data release of the LOFAR LBA Sky Survey, including source extensions, sensitivity, astrometric accuracy and precision, and flux density uncertainties. 

\subsection{Source sizes}
\label{sec:sourcesize}

Separating point-like from extended sources is notoriously difficult in regimes where phase errors are an important systematic effect. A perfectly point-like source, in the absence of calibration and deconvolution errors, is defined to have a ratio of integrated flux density ($S_I$) to peak flux density ($S_P$) equal to unity, with a source size equal to that of the restoring beam. As discussed in Sec.\ 3.1 of \cite{Shimwell2022}, in optimal conditions the natural logarithm of the ratio of the two quantities, $R=\ln(S_I/S_P)$, is expected to have a Gaussian distribution in absence of extended sources. However, even for a distribution of point-like sources (i) the non-Gaussian rms noise of the map, (ii) the correlation between the error of the two quantities, and (iii) the increase of errors and general overestimation of the source sizes as $S_I$ decreases (for instance due to incomplete deconvolution), all combine to skew the distribution of $R$. The actual distribution is further complicated by the quantisation of the pixels in our images and the residual time- and bandwidth-smearing effects. Most importantly, any uncorrected variation in the ionospheric-induced phase shift adds to the source smearing, therefore artificially increasing the ratio $R$.

\begin{figure}
\centering
 \includegraphics[width=.5\textwidth]{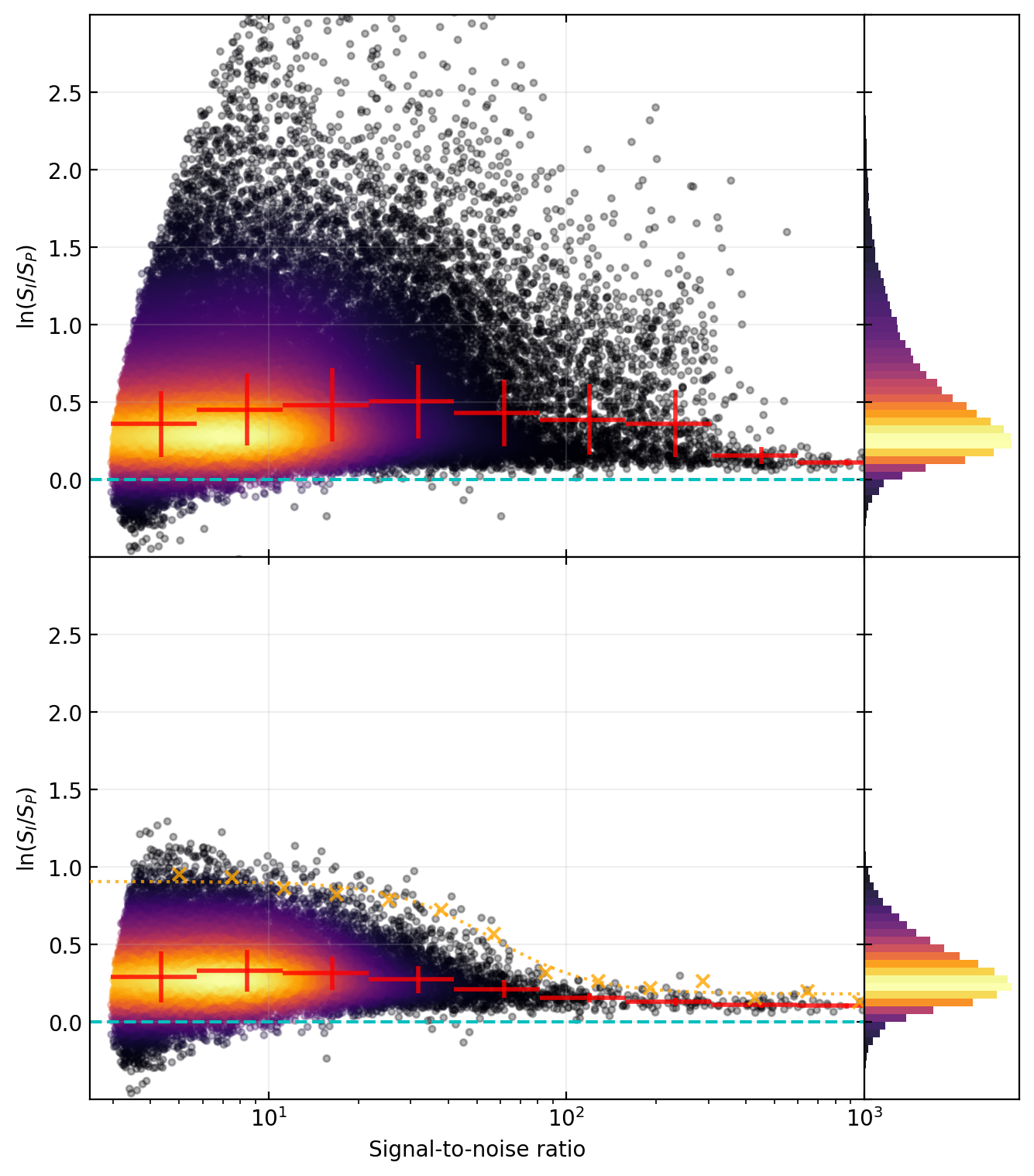}
 \caption{The logarithm of the ratio of integrated to peak flux density plotted against the signal-to-noise ratio as measured by the source finder. A perfect point source would have $\ln(S_I/S_P) = 0$ (blue dashed line). In the top panel we show all isolated sources present in the catalogue. The bottom panel is restricted to those sources whose cross-match is an isolated point source in LoTSS (see text). Red crosses show the binned median with the extension on the x-axis showing the bin size and on the y-axis showing $\pm 1$ median absolute deviation. Orange markers show the position of the 99th percentile of the distribution in each bin. Dotted orange line is a fit to the orange markers positions as explained in the text.}
 \label{fig:inttopeak}
\end{figure}

In order to decide whether a source is extended in LoLSS, we adopted a strategy similar to the one used for LoTSS, but taking advantage of the information obtained by combining the two surveys. Firstly, we located all isolated sources in the overlapping region of LoTSS DR2 and LoLSS, where ``isolated" means that no other source in the same survey was detected within 30\arcsec. Isolated sources of LoTSS and LoLSS are then cross-matched using a maximum distance of 15\arcsec. The LoLSS signal-to-noise ratio (SNR; defined as for LoTSS with $S_I/\sigma_{S_I}$) of isolated sources is plotted on the top panel of Fig.~\ref{fig:inttopeak} against $R=\ln(S_I/S_P)$. It is evident that all sources are affected by smearing, most likely dominated by residual ionospheric-induced phase errors combined with the other aforementioned effects.

In the bottom panel of Fig.~\ref{fig:inttopeak}, we restricted the catalogue to those sources classified as point sources in LoTSS following the definition of Sec.~3.1 of \cite{Shimwell2022}. This reduces the number of sources by about a factor of two. About 4\% of sources were also removed because they are composed by multiple gaussians or the size of the fitted gaussian is larger than 30\arcsec. Given the higher sensitivity and resolution of LoTSS, this procedure should ensure that the selected sources are point-like in LoLSS. We then binned the sources in ten SNR ranges starting from SNR~$=4$. The position of the 99th percentile was calculated in each bin and a sigmoid function was used to fit the envelope, finding:

\begin{equation}\label{eq:sigmoid}
R_{95} = 0.18 + \left( \frac{0.73}{1+\left( \frac{\rm SNR}{56.00} \right)^{2.56}} \right).
\end{equation}

This function can be used to separate point from resolved sources in the LoLSS catalogue. This approach is more accurate than the simple $S_I/S_P$ ratio or the use of fitted Gaussian sizes because it takes into account the effect of the SNR. At low SNR sources can be only partially deconvolved, due to non-detection in some of the mosaiced images, and/or smeared by approximate ionospheric corrections when they are far away from bright direction-dependent calibrators. By applying our criterion, we find that 9170 of LoLSS sources (21.8\%) are classified as extended.

\subsection{Sensitivity}
\label{sec:sensitivity}

\begin{figure*}[htb]
\centering
 \includegraphics[width=\textwidth]{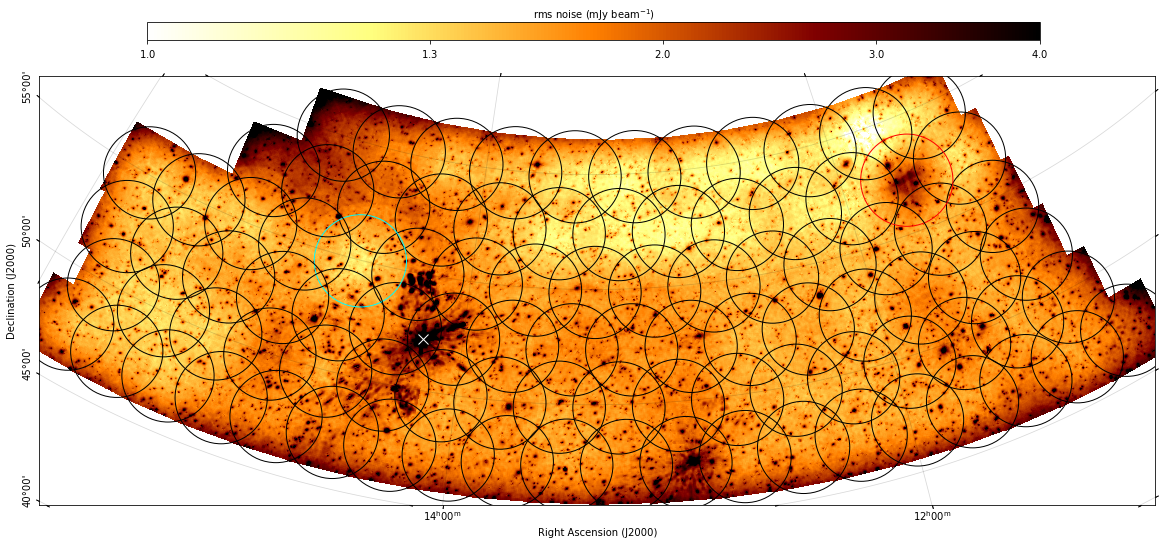}
 \caption{Rms noise map of the LOFAR LBA Sky Survey - DR1. The regions with reduced sensitivity are located around bright sources that induce dynamic range limitations. Each pointing is shown with a circle at the FWHM. The red pointing (P174+57) was not observed, while the blue pointing (P218+55) was observed for 16 hrs (twice as much as the others). The white ``x'' indicates the position of the bright source 3C\,295, which lower the dynamic range in several fields around it.}
 \label{fig:rmsmap}
\end{figure*}

In Fig.~\ref{fig:rmsmap} we present the spatial distribution of the local rms noise value across the survey area. This was derived using the PyBDSF rms noise estimation. Rms noise images from each $3.3\deg\times3.3\deg$ mosaic were reprojected and combined into a single large mosaic using montage\footnote{\url{http://montage.ipac.caltech.edu/}.}. The main source of local variation in the rms noise map are dynamic range limitations close to bright sources. The clearest example is seen around 3C\,295. that creates the large low-sensitivity region visible in Fig.~\ref{fig:rmsmap} around its position, marked with an ``x''. Observations for the two northern strips were taken in 2019, while the rest of the region was covered in 2017, when the solar activity was higher due to the phase of the 11-yr solar cycle. A high solar activity induces strong ionospheric disturbances that result in more severe systematic errors in the data. While most of the fields include 7 to 8 hours of good data, P218+55 (marked in blue in Fig.~\ref{fig:rmsmap}) was observed for 16 hrs, and hence shows a better local sensitivity. On the other hand, field P174+57 (marked in red in Fig.~\ref{fig:rmsmap}) was not observed and the region was covered only by neighbouring fields resulting in a reduced local sensitivity.  

\begin{figure}
\centering
 \includegraphics[width=.5\textwidth]{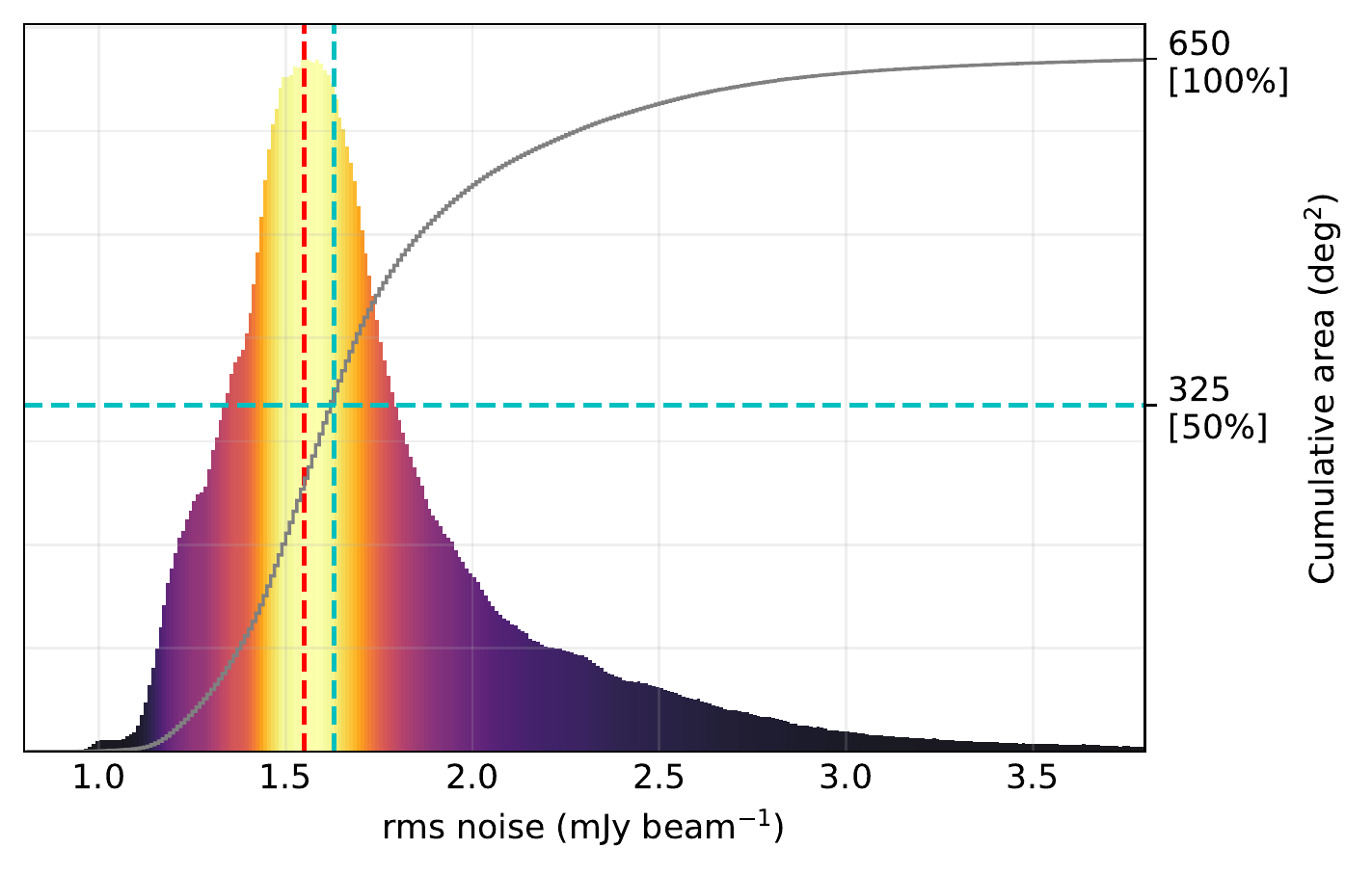}
 \caption{Rms noise histogram of the pixels included in the released region. The solid line shows the cumulative function. The red dashed line indicates the the rms value at the peak of the distribution (1.55~\mjybeam). The blue dashed line shows the position of the 50\% percentile (median) at $1.63$~\mjybeam. This means that half of the covered area (325 deg$^2$) has a lower rms noise than that. The long tail of high rms noise values is due to bright sources in the field.} \label{fig:rmshist}
\end{figure}

In Fig.~\ref{fig:rmshist} we show the distribution of the pixel values of the rms mosaic shown in Fig.~\ref{fig:rmsmap}. The distribution is not symmetric due to a long tail of high rms noise pixels caused by dynamic range limitations around bright sources and edge effects. The peak of the distribution at an rms noise of 1.55~\mjybeam. This value can be considered representative for the data quality of LoLSS. The median value is 1.63~\mjybeam{}.

\begin{figure}
\centering
 \includegraphics[width=.5\textwidth]{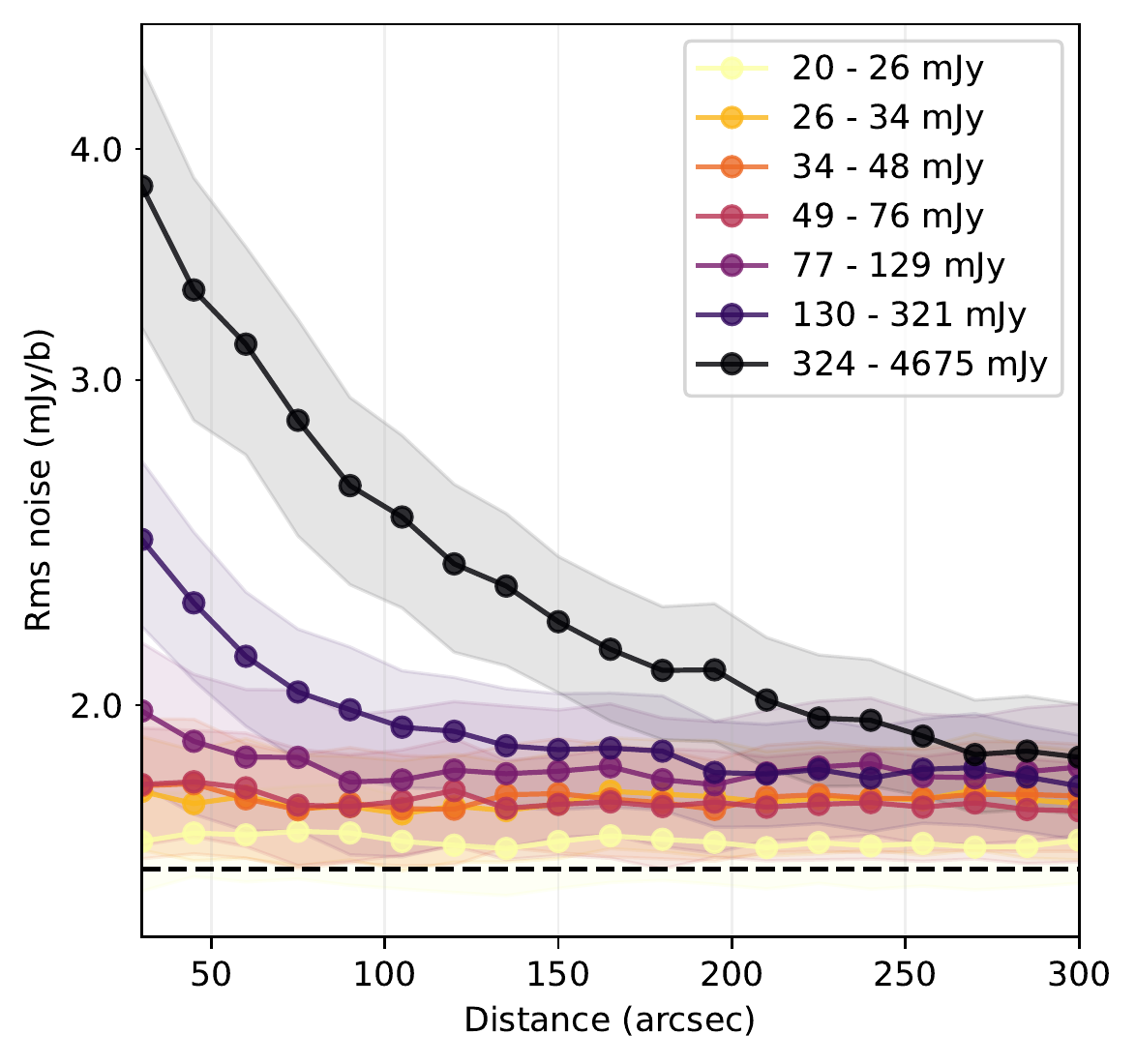}
 \caption{Dynamic range limitation around bright sources. The $x$-axis shows the distance in arcsec from bright isolated sources. The $y$-axis shows the local rms noise. Each line represents the median for about 190 sources in different flux density intervals with a shaded region that indicates one median absolute deviation. The black dashed line is at 1.63~\mjybeam.}
 \label{fig:dynrange}
\end{figure}

In order to estimate the effect of dynamic range limitations around bright sources we located all sources that are clearly isolated, i.e. with no other detected sources within 450\arcsec{}. Then we estimated the rms noise using the residual mosaic maps in concentric annuli with a thickness of 15\arcsec{}. Sources were then binned based on their flux density and the median rms noise was estimated as a function of distance from the source position (see Fig.~\ref{fig:dynrange}). The procedure shows an increase in the rms noise close to sources brighter than $\sim 100$ mJy.

\subsection{Astrometric precision and accuracy}
\label{sec:astrometry}

Before performing mosaicing of individual images we applied an astrometric correction based on the offset with cross-matched FIRST sources. To estimate the global astrometric precision and accuracy, we selected point-like (as defined by eq.~\ref{eq:sigmoid}) and isolated (no other source within 45\arcsec) LoLSS sources, as well as all isolated FIRST sources (no other source within 15\arcsec). We then cross-matched the resulting catalogues. The cross-match was done starting with a maximum separation of 100\arcsec{} and keeping only sources within 10 times the median absolute deviation (MAD) of all distances between matched sources. This procedure is then applied iteratively, recalculating the median absolute deviation until it converges. The final matching distance is 6\arcsec{}, which gives 12\,375 LoLSS sources with an associated FIRST counterpart.

\begin{figure}
\centering
 \includegraphics[width=.5\textwidth]{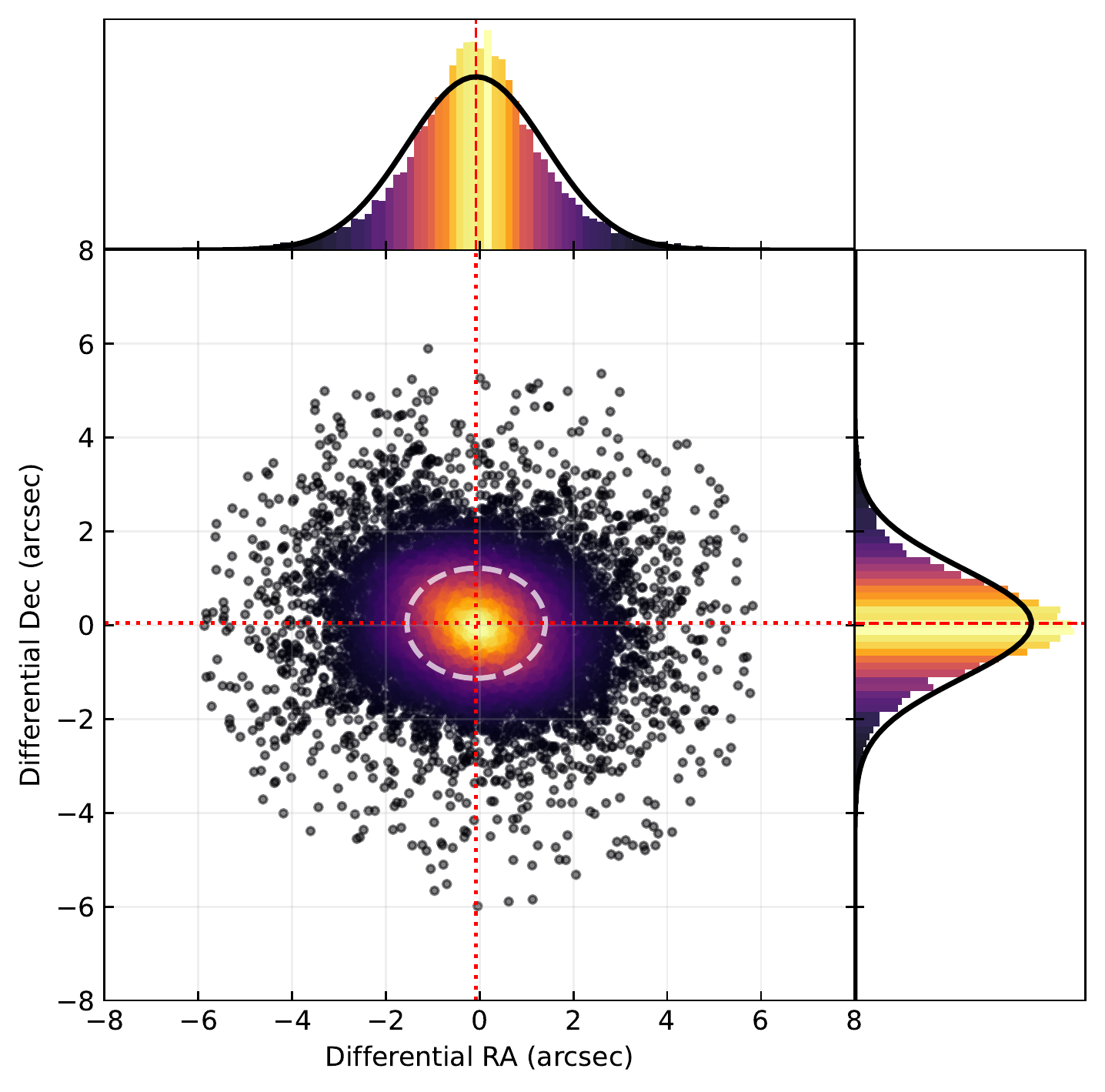}
 \caption{Astrometric accuracy of the sources present in the catalogue measured using the FIRST survey as a reference. The mean errors are $E_{\rm RA} = -0.08\arcsec$ and $E_{\rm Dec} = 0.04\arcsec$ with standard deviation $\sigma_{\rm RA} = 1.48\arcsec$ and $\sigma_{\rm Dec} = 1.17\arcsec$. The standard deviation of the distribution is shown with a white dashed line. In the histograms, the black solid lines show Gaussians with the aforementioned means and standard deviations, the red dashed lines indicate the position of the means.}
 \label{fig:astrometry}
\end{figure}

In Fig.~\ref{fig:astrometry} we plot the RA and Dec separation between the matched sources. We can then use the mean separations $E_{\rm RA} = -0.07\arcsec$ and $E_{\rm Dec} = 0.04\arcsec$ as an estimate for the astrometric accuracy, and the standard deviation $\sigma_{\rm RA} = 1.44\arcsec$ and $\sigma_{\rm Dec} = 1.13\arcsec$ as an estimate of the astrometric precision.

\subsection{Flux density uncertainties}
\label{sec:flux}

The major source of uncertainties in the LoLSS flux density scale is the analytic LOFAR beam model that is used to compensate for the flux density suppression due to both the dipole beam and the station beam. For LoLSS observations, the instrumental bandpass was estimated using the calibrator beam and a calibrator model outlined by \citet{Scaife2012}. The bandpass solutions were then transferred to the target beams so that the flux density scale matches the one used for the calibrator model. No further corrections were applied to the flux density scale, which makes LoLSS independent of other radio surveys. Therefore, the accuracy is limited only by instrumental stability and the primary beam model. This is in contrast to LoTSS, where the 6C and NVSS surveys were used to re-scale the LoTSS images to the expected flux density scale \citep[][Sec. 3.3.1]{Shimwell2022}.

Following \cite{deGasperin2021}, we can give an initial estimate of the expected flux scale uncertainty. The first source of uncertainty is due to the \cite{Scaife2012} flux density scale, that has a nominal error ranging between two and four percent depending on the calibrator. The second uncertainty is due to the dipole beam model errors that in a mosaic is dominant over the uncertainties of the primary beam model far from the phase centre. These were estimated by observing two calibrators simultaneously and comparing the derived bandpass solutions. This experiment showed an error of about five percent. Finally, an estimate of the flux density uncertainty due to calibration and imaging processes was derived by measuring the flux density of the calibrator source 3C\,295, which is within the survey footprint. The measured flux density of that source is 134.4 Jy compared to an expected flux density of 133.3 Jy, a difference of one percent. Adding the three uncertainties in quadrature results in a combined systematic uncertainty of 6 percent.

In \cite{deGasperin2021}, we emphasised the presence of a 10\% systematic discrepancy between the flux density scale of LoLSS and LoTSS. However, the discrepancy was derived using images without direction-dependent corrections and assuming a straight extrapolation of the spectral index from NVSS and LoTSS, down to LoLSS frequencies. Consequently, it was unclear how much of this effect could be attributed to the average spectral curvature of radio sources. With the improved data of LoLSS DR1, combined with the in-band spectra, this can be evaluated in more detail.

For this purpose, we used only isolated detections (i.e. no other sources within twice the survey resolution with a minimum of 30\arcsec) and we cross-matched all surveys with LoLSS using a maximum distance for a positive match of 5\arcsec{} for LoTSS, 10\arcsec{} for NVSS, 20\arcsec for VLSSr, and 30\arcsec{} for 8C. To avoid confusion due to the low resolution of the 8c survey, we discarded any matched LoLSS source whose flux density accounts for $<90\%$ of the sum of the flux densities of the LoLSS sources present in the 8c beam. Also all sources composed by multiple gaussians and with a signal-to-noise ratio $<10$ are removed. This reduces the number of LoLSS sources to 11\,810. Each of these sources may or may not have one or more cross-matches in other surveys.

\begin{figure}
\centering
 \includegraphics[width=.5\textwidth]{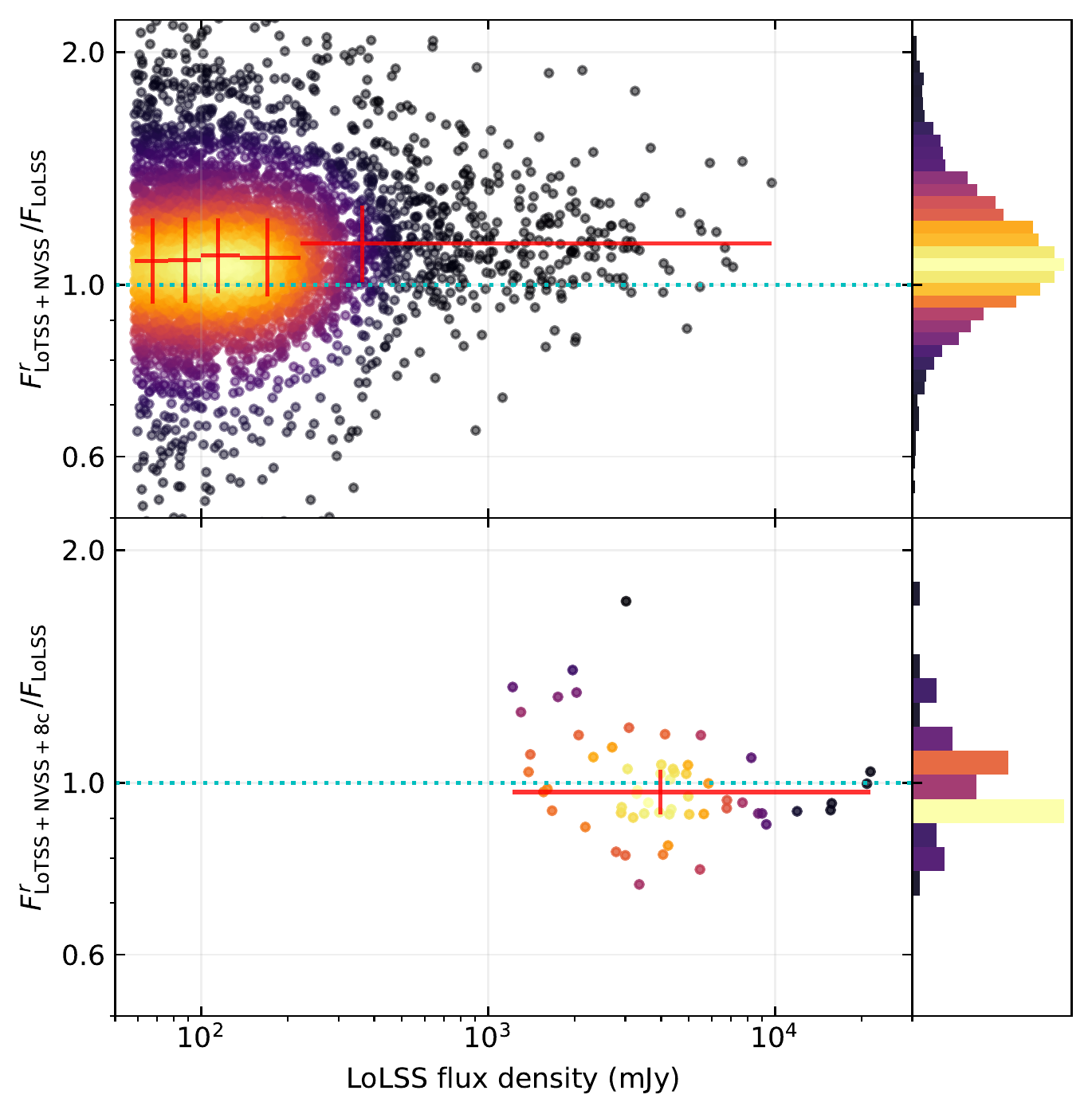}
 \caption{LoLSS flux density versus the expected flux density at 54 MHz from the linear extrapolation of the spectral energy distribution (SED) using NVSS and LoTSS (top panel) or from a quadratic polynomial SED estimated using NVSS, LoTSS, and 8C (bottom panel). A ratio of 1, the blue dotted lines, means a perfect match between the prediction and the flux density measured in LoLSS. Red crosses are binned medians (five bins for the top panel, one for the bottom) with $\pm 1$ median absolute deviation shown as an extension in the y-direction and the bin size as an extension in the x-direction.}
 \label{fig:fluxexp}
\end{figure}

Firstly, we compared the measured LoLSS flux density with the expected value at 54 MHz estimated via a linear extrapolation of the spectral energy distribution (SED) that is derived from NVSS and LoTSS flux densities. We reduced the number of usable sources to 4\,895 by applying a flux-limit on each survey so that every source with a spectral index $-0.5<\alpha<-1.0$ was catalogued in all surveys. This constraint should reduce the bias due to the different flux density cuts of the various surveys. The results are shown in the top panel of Fig.~\ref{fig:fluxexp}. In this case, the flux density of LoLSS is lower than predicted by $7-13$ percent, depending on the flux density (median: $1.09\pm0.01$; MAD: 0.13). We note that regions of the survey with higher rms noise correspond to higher flux density deficit, this is most likely related to ionospheric smearing combined with our inability to make a complete correction of the effect. In the lower panel of the same figure, we compared the LoLSS flux density with the expected value at 54 MHz derived from a second order polynomial, therefore a curved spectrum, obtained using NVSS, LoTSS, and 8C. Although 8C has a partial overlap with LoLSS footprint and a low sensitivity and resolution ($\sim 200$~\mjybeam with a beam size of \beam{270}{270}), the survey is at 38~MHz, a lower frequency than LoLSS, which allows a better constraint compared with the LoTSS -- NVSS extrapolation. In this case, we have only 61 matched sources but the flux density of LoLSS is well aligned with predictions with a median separation that is two percent (median: $0.97\pm0.05$; MAD: 0.06).

\begin{figure}
\centering
 \includegraphics[width=.5\textwidth]{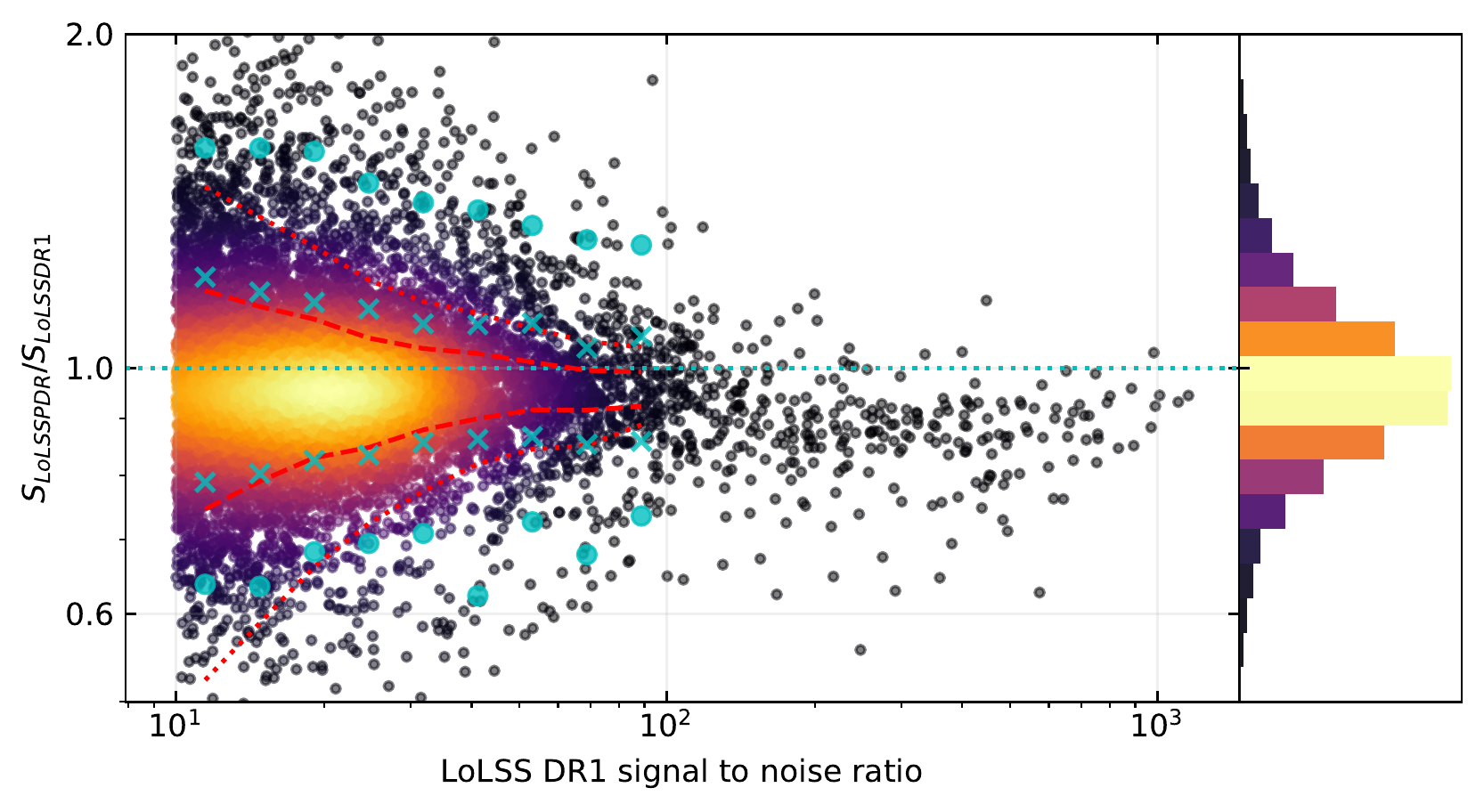}
 \caption{Ratio of the flux densities of matched sources in LoLSS PDR and DR1 as a function of the signal-to-noise ratio in LoLSS DR1. Red dashed and dotted lines are the expectation for the $1\sigma$ and $2\sigma$ dispersion due to the rms noise in the two surveys, cyan crosses and circles are the $1\sigma$ and $2\sigma$ points evaluated in various bins showing larger wings compared to expectations.}
 \label{fig:comparisonpdr}
\end{figure}

We compared the flux densities of sources in LoLSS First Data Release (LoLSS DR1) with the flux densities from the Preliminary Data Release (LoLSS PDR). We show the comparison in Fig.~\ref{fig:comparisonpdr}, here the ratio of the flux densities of matched sources from the two releases are plotted against the signal-to-noise ratio in DR1. The median deviation between the two surveys is 4\% (MAD: 0.1). For each matched source, the errors in the flux density estimations are loosely correlated as most of the uncertainty comes from a combination of systematic effects that are treated differently in the two releases. To check the precision of the surveys we can derive the expected distribution of their flux density ratios. This is done extracting the flux density of each source twice, once for the PDR and once for DR1, and taking the ratio. For each source, the two flux densities are estimated taking the flux density value reported in DR1 and modifying it by adding a value extracted from a normal distribution with standard deviation equal to the flux density uncertainty as catalogued in the two releases. The distribution of these flux density ratios is compared with the real one in Fig.~\ref{fig:comparisonpdr}. The data match the expectations for the most part, showing that the precision of the flux density is dominated by the survey rms noise. However, it is evident that the number of outliers is larger compared to expectations. This might be due to systematic effects dominating over the noise in the PDR.

These results are in line with most of LoLSS sources having slightly curved spectra with a possible reduced flux density at 54 MHz compared with a linear extrapolation from higher frequencies. However, the flux density scales of LoLSS and LoTSS systematic uncertainties are compatible with a straight average spectrum. We conclude that LoLSS does not possess an appreciable systematic flux scale offset. The precision of the LoLSS flux density scale is estimated to be 6 percent.

\subsection{In-band spectral index}
\label{sec:in-band}

\begin{figure*}[htb!]
\centering
 \includegraphics[width=.33\textwidth]{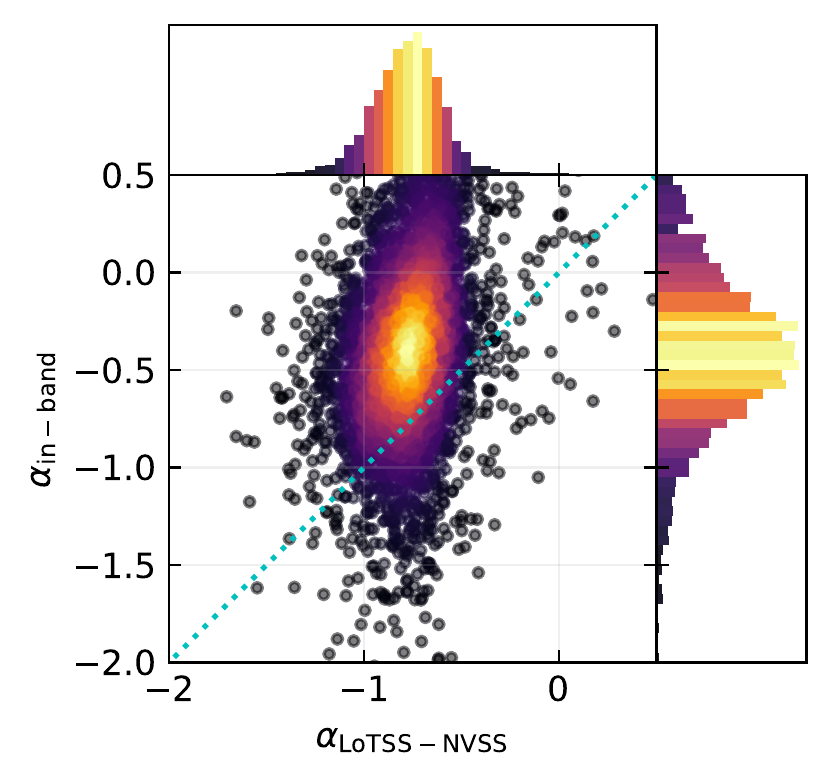}
 \includegraphics[width=.33\textwidth]{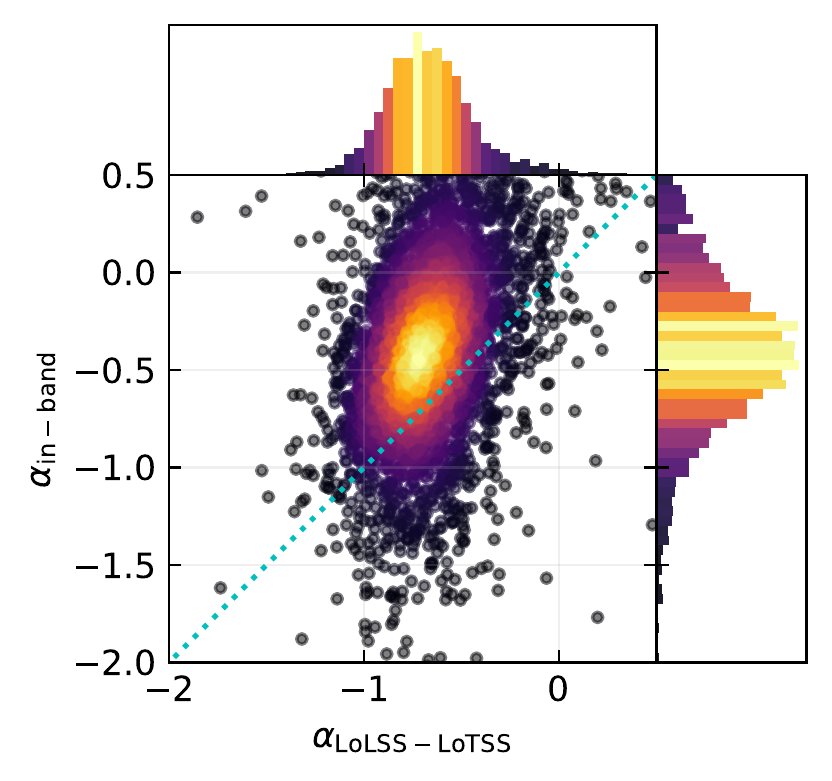}
 \includegraphics[width=.33\textwidth]{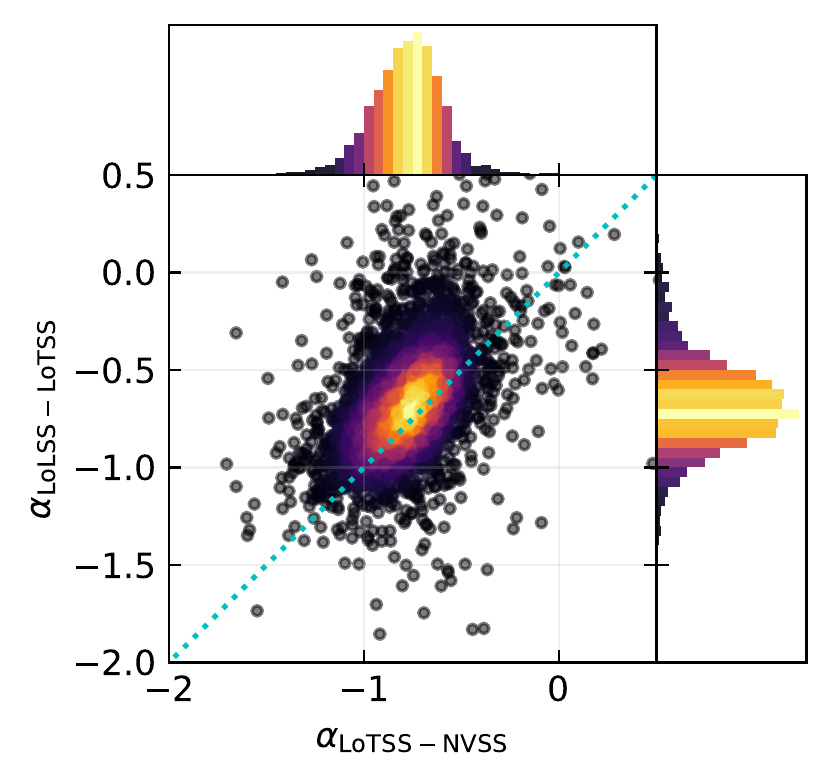}
 \caption{Left: high-frequency wide-band radio spectral index ($144-1400$ MHz) from LoTSS and NVSS versus the LoLSS in-band spectral index ($42-66$ MHz). Center: the same but using a low-frequency wide-band radio spectral index ($54-144$ MHz) from LoLSS and LoTSS. Right: comparison between the high- and the low-frequency wide-band spectral index.}
 \label{fig:inbandspectra}
\end{figure*}

The in-band spectral indices of LoLSS were compared with the wide-band spectral indices as derived from other surveys. This comparison is presented in the first two panels of Fig.~\ref{fig:inbandspectra}, where LoLSS in-band spectral indices are compared with the wide-band spectral indices at high frequency (LoTSS -- NVSS) and low frequency (LoLSS -- LoTSS). In both cases, the in-band spectral indices are flatter than the wide-band spectral indexes with a median in-band spectral index of $-0.37\pm0.01$ (MAD: 0.27) compared with a median wide-band spectral index of $\alpha_{\rm LoTSS-NVSS} = -0.769\pm0.004$ (MAD: 0.11) and $\alpha_{\rm LoLSS-LoTSS} = -0.680\pm0.005$ (MAD: 0.14). Using only half of the sources with the highest signal-to-noise ratio does not change the result appreciably. In the last panel of Fig.~\ref{fig:inbandspectra}, we compare the two wide-band spectral indices. This last panel shows a median flattening of the spectral index of $-0.091\pm0.005$ moving towards the lower frequencies. This, however, is clearly smaller than the large flattening shown by the in-band spectra, pointing towards a possible systematic error in the latter.

\begin{figure}[htb!]
\centering
 \includegraphics[width=.5\textwidth]{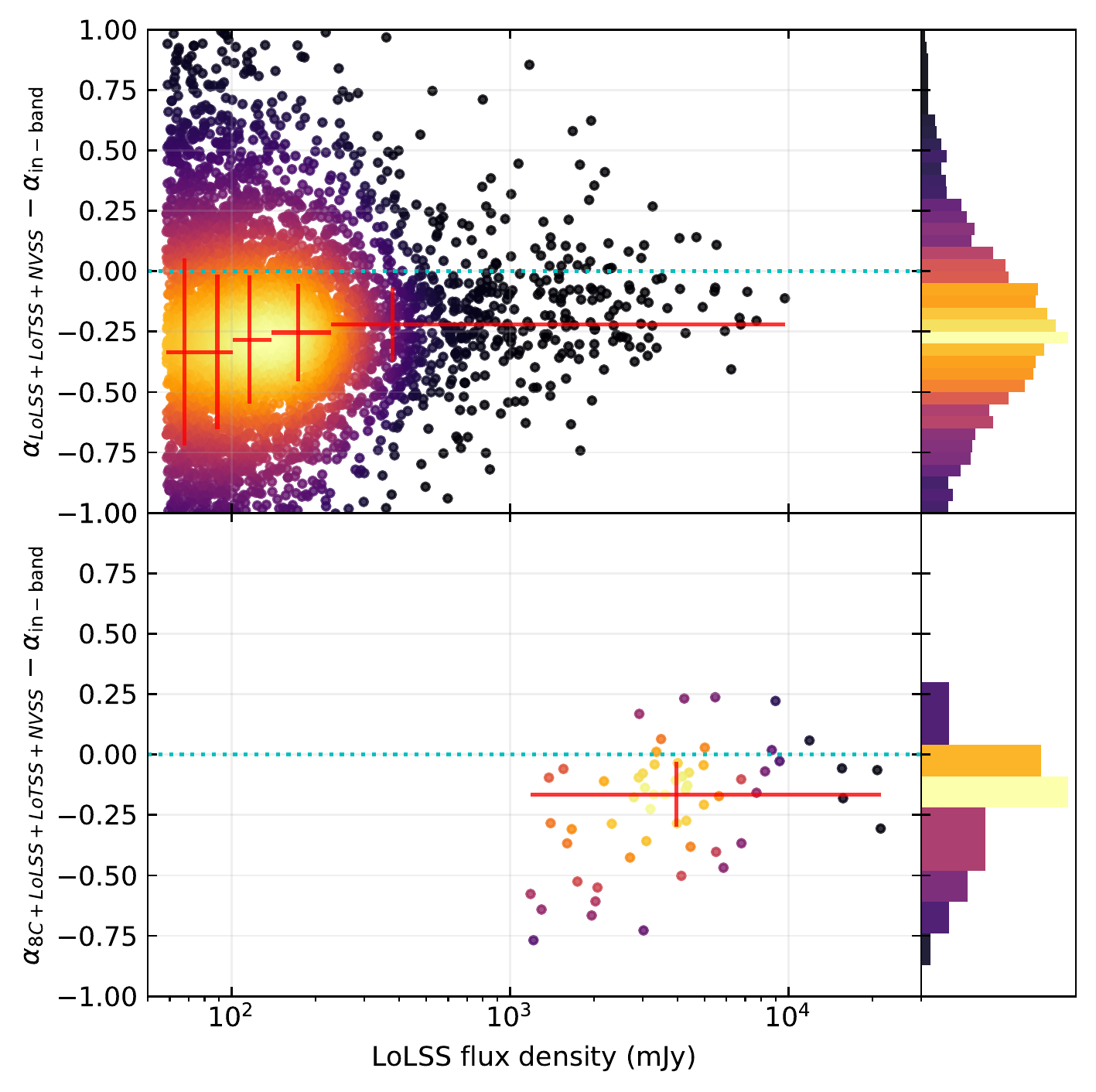}
 \includegraphics[width=.5\textwidth]{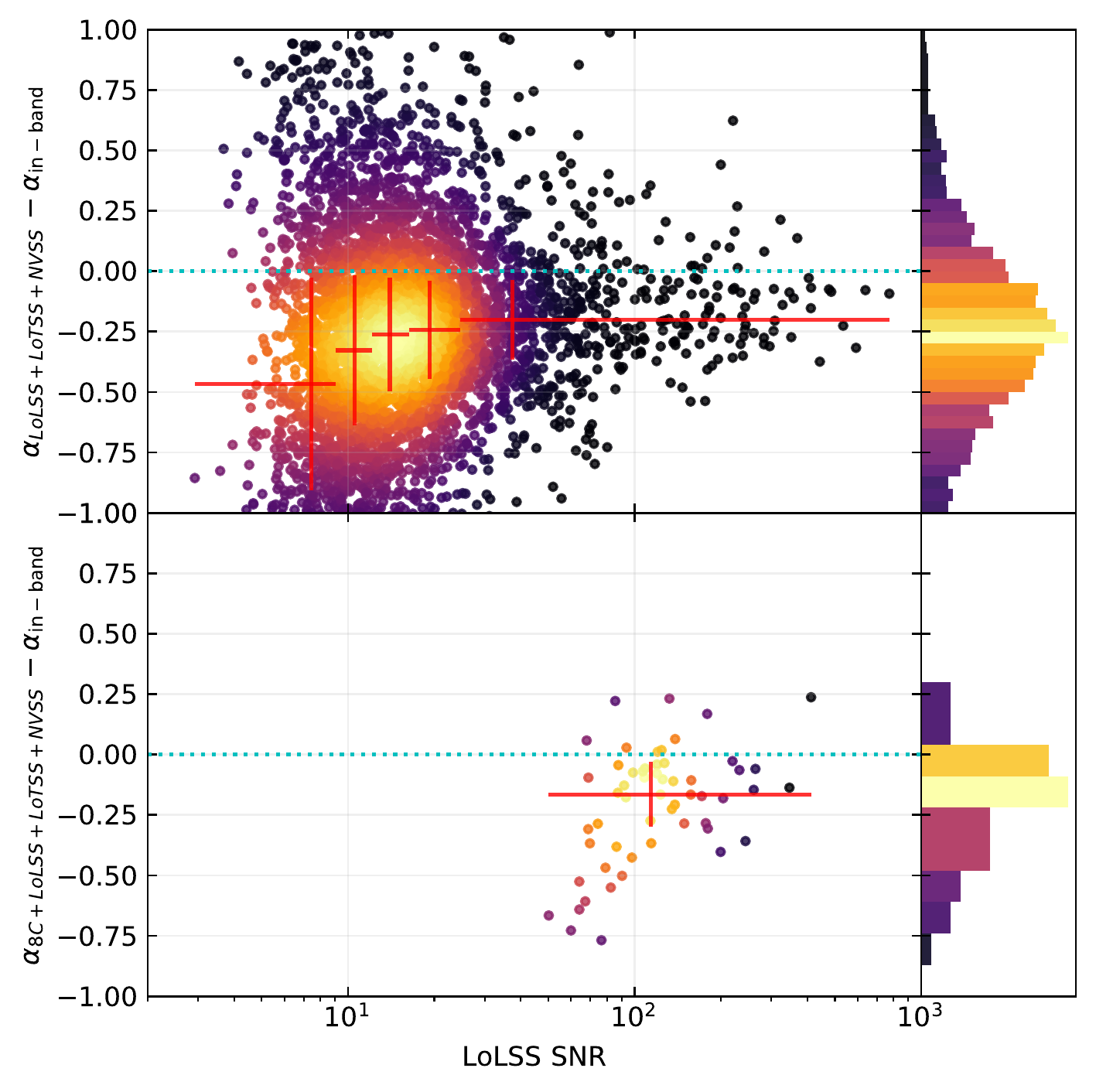}
 \caption{In-band spectral index separation with the tangent of a second order polynomial fit to the SED derived using LoLSS, LoTSS, and NVSS (top panel of each figure) and 8C, LoLSS, LoTSS, and NVSS (bottom panel of each figure). The two figures show the spectral index difference as a function of flux density (top) and signal-to-noise ratio (bottom). Red crosses are binned medians with $\pm 1$ median absolute deviation showed as an extension in the y-direction and the bin size as an extension in the x-direction. The average in-band spectral index results flatter than expectations.}
 \label{fig:spidxinbandexp}
\end{figure}

The comparisons done so far are subject to effects due to the different frequency ranges used for the in-band spectral indices and the possible curvature of the source SEDs. To further investigate to what extent the in-band spectra are reliable, we fitted a quadratic polynomial to the flux density values of LoLSS, LoTSS, and NVSS. The tangent to these polynomial, evaluated at 54 MHz is then compared with the in-band spectral index (see Fig.~\ref{fig:spidxinbandexp}). Again the in-band spectral index are flatter than expectations, with a median difference between expectation and in-band of $-0.27\pm0.01$ (MAD: 0.25). The median becomes $-0.17\pm0.07$ (MAD: 0.13) when considering also 8C although this reduces the sample to just 58 sources (one not being detected in all the in-band channels). Our conclusion is that a certain degree of flattening is expected and encoded in the in-band spectral index; however, a systematic offset of between $+0.2$ and $+0.3$ might be present. This is likely due to an incomplete model of the LOFAR LBA element beam. We also note that for sources with low signal-to-noise ratio the systematic offset might increase to $+0.4$ (see Fig.~\ref{fig:spidxinbandexp}). Another complexity can come from the independent amounts of Eddington bias \citep{Eddington1913}, which brightens sources close to the noise threshold that varies in each band and with the sky position. Finally, we checked that the average in-band flux density is compatible with the one reported in the main catalogue and we find a median deviation $<0.2\%$.

\begin{figure}
\centering
 \includegraphics[width=.5\textwidth]{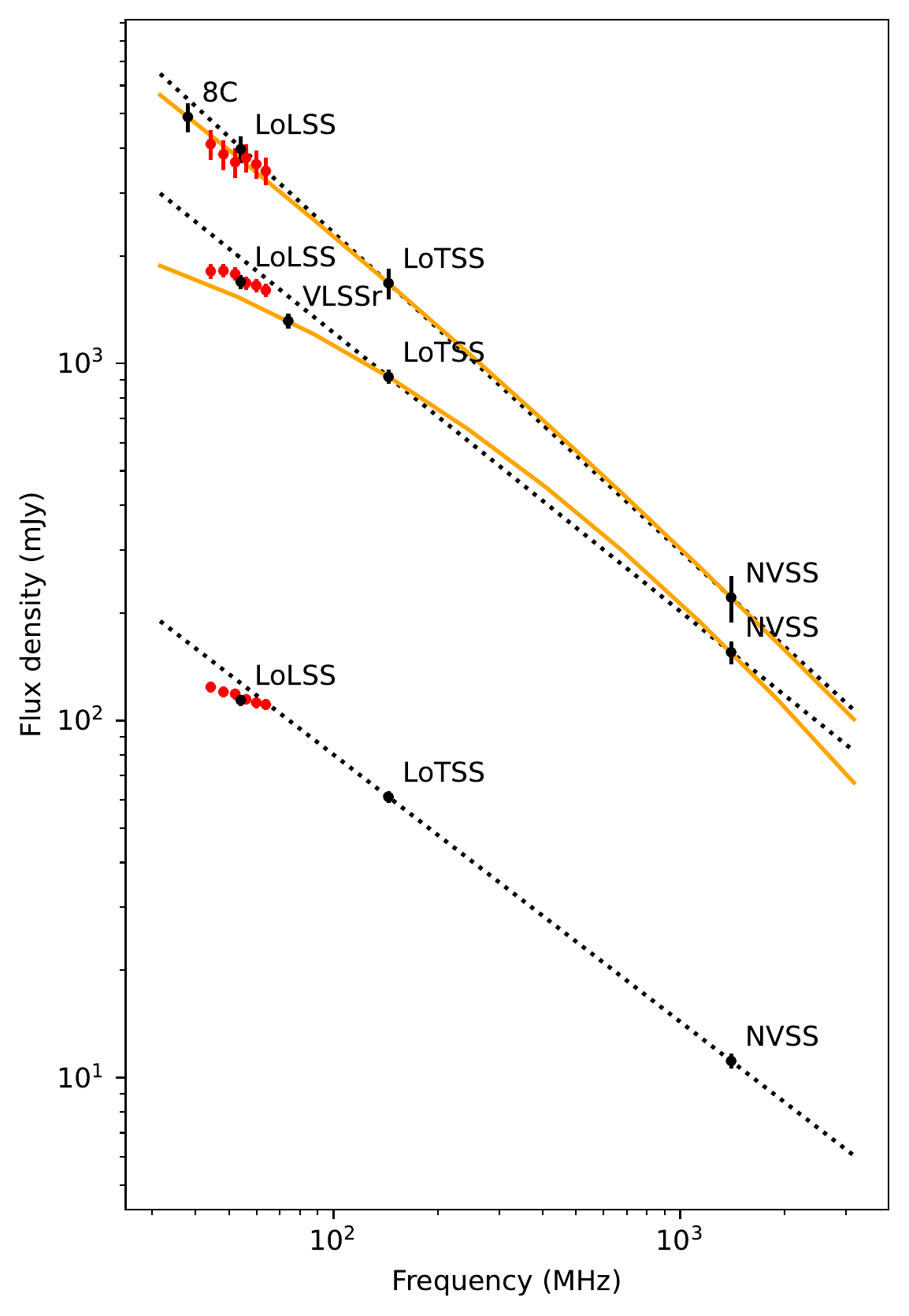}
 \caption{Median flux densities of all matched sources for 8C, LoLSS, LoTSS, and NVSS (61 sources; top line), LoLSS, VLSSr, LoTSS, and NVSS (130 sources; mid line) and LoLSS, LoTSS, and NVSS (4\,895 sources; bottom line). In red are the LoLSS in-band flux densities. The height of the lines shows the uncertainty on the median. Black dotted lines are linear polynomials derived from NVSS and LoTSS values in all cases. Yellow line is the quadratic polynomial derived from 8C/VLSSr, LoTSS, and NVSS.}
 \label{fig:singlespectrum}
\end{figure}

In Fig.~\ref{fig:singlespectrum}, we show the median flux density value of all matched sources for three sets of radio surveys. The plot underlines the overall good alignment of the various flux scales as well as a local flattening at low frequencies. The flattening is also present in the in-band spectral index but, as discussed above, stronger than predicted. 

\begin{figure*}
\centering
 \includegraphics[width=\textwidth]{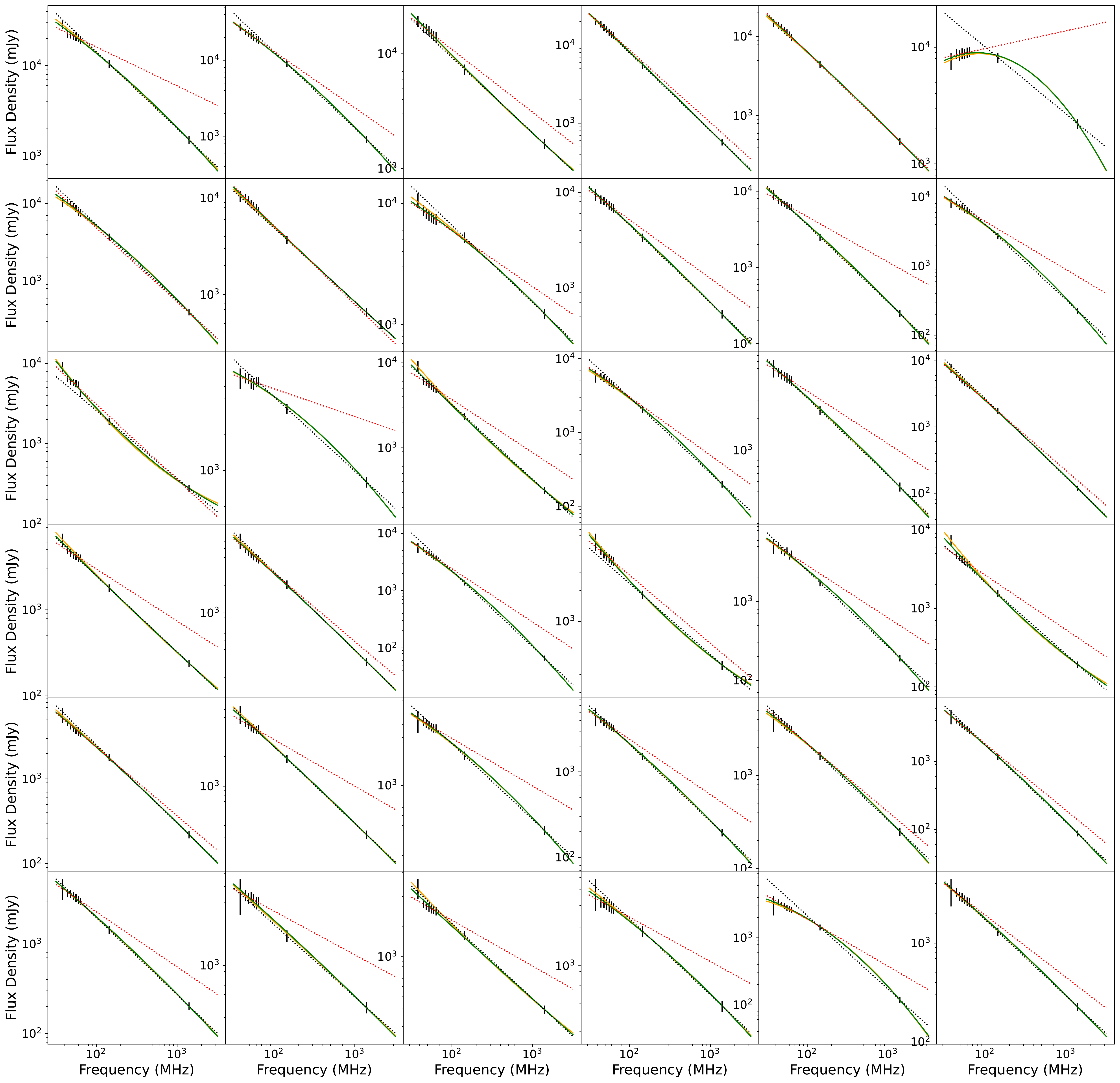}
 \caption{Examples of radio SEDs of isolated bright sources with a detected 8C counterpart. In each plot, the black vertical lines show the data points of 8C (38 MHz), LoLSS (44, 48,  52, 56,  60, 64 MHz), LoTSS (144 MHz) and NVSS (1400 MHz) with uncertainties. The black dotted line connects LoTSS and NVSS data points. The red dotted line is a linear regression using only the LoLSS in-band data points. The green line is a 2nd order regression including all data points while the yellow line is a 2nd order regression without considering LoLSS data points. The green and yellow lines are almost always superimposed. }
 \label{fig:spectra}
\end{figure*}

Finally, we examined some of the brightest sources for which 8C, LoLSS, LoTSS, and NVSS flux densities are positively matched. For these sources we show in Fig.~\ref{fig:spectra} the extrapolated linear spectra using LoTSS and NVSS (yellow dotted lines) and the second order polynomial fit done using 8C, LoTSS, and NVSS (green line) and including also in-band LoLSS values (red line). In all cases, LoLSS data seem to be fit well with a second order polynomial function although in certain cases the in-band spectra index (black dotted lines) results are flatter than expected. The shape of the functions differ only marginally when LoLSS data are included in the fit, showing again a good overall agreement.

\section{Discussion}

\subsection{Curved spectra}

The two main LOFAR surveys, LoTSS (144 MHz) and LoLSS (54 MHz) are highly complementary, with the first having higher resolution and sensitivity and the second providing information on the low-frequency spectral index. Currently, no other existing or planned instrument besides LOFAR LBA will be able to deliver information on the radio SED down to tens of MHz, where only the brightest radio sources have been studied up till now.

The catalogue presented in this work showed that the average radio SED shows a curvature at frequencies between $50 - 1400$~MHz, going from a median high-frequency spectral index $\alpha_{\rm LoTSS-NVSS} = -0.769\pm0.004$ (MAD: 0.11) to a median low-frequency spectral index $\alpha_{\rm LoLSS-LoTSS} = -0.680\pm0.005$ (MAD: 0.14), see also Boehme et al. (in prep.) where LoLSS preliminary release sources matched to other radio surveys also hint at curved spectra. Other low-frequency radio surveys, such as GLEAM, have found a relatively small fraction ($<10\%$ percent) of curved spectra \citep{Callingham2017}.
%It is important to note that GLEAM reaches down to 72 MHz, making it slightly harder to identify a low-frequency deviation from a straight spectrum.

A deviation from straight spectra is not new for samples of sources observed at low-frequencies. \cite{Laing1980}, for instance, showed that for a sample of sources selected at 178 MHz, the deviation of the measured spectra from the flux density, predicted with a power-law fit, increased with decreasing frequency. This was an indication of curved spectra becoming increasingly dominant at frequencies $<100$~MHz. However, their analysis was restricted to sources of type II in the Fanaroff-Riley classification scheme \citep[FR\,II sources;][]{Fanaroff1974} whose flux density is strongly affected or dominated by the emission from their hot-spots. Consequently, their interpretation was that self-absorption in the compact hot-spots was the main driver for the downward curvature \citep{Laing1980}.

For the first time, our catalogue provides flux density measurements at 54~MHz for a large population of radio sources. At the few mJy level reached by LoLSS, the dominant population are FR\,I radio galaxies \citep{Mingo2019}. It is important to note that star-forming galaxies, well represented in the LoTSS survey catalogue, are subdominant in LoLSS and the detection of this population at 54 MHz requires a sensitivity of $0.5-1$ mJy \citep{Williams2021}.

Unlike FR\,II radio galaxies, at low frequencies FR\,I radio galaxies are not dominated by the emission of compact regions, but rather by the diffuse emission of their lobes. This makes any absorption mechanisms less likely to cause the turnover. A possible explanation for low-frequency-curved spectra in this source class is that integrated radio SED are a superimposition of aged spectra with a rather flat ($\alpha \sim -0.5$) injection index and the ultra-low frequencies are dominated by emission from low-energy electrons that have not lost most of their energy yet. Therefore, their energy distribution generates radio emission that preserves a flatter spectral index, closer to the injection index. Alternatively, we might be witnessing the beginning of the low-energy cutoff of the electron distribution. Particle acceleration processes in radio AGN are expected to accelerate electrons following a power-law in momentum giving, at relativistic particle energies, an energy distribution $n(E)dE \propto E^{-p} dE$, which translates into a straight radio SED with spectral index $\alpha = \frac{1 - p}{2}$. However, the number of electrons at low energies cannot increase indefinitely and a cutoff is required. A detailed analysis of this result is outside of the scope of this paper and will be presented in a forthcoming publication.

\subsection{Source counts}

\begin{figure}
\centering
 \includegraphics[width=.5\textwidth]{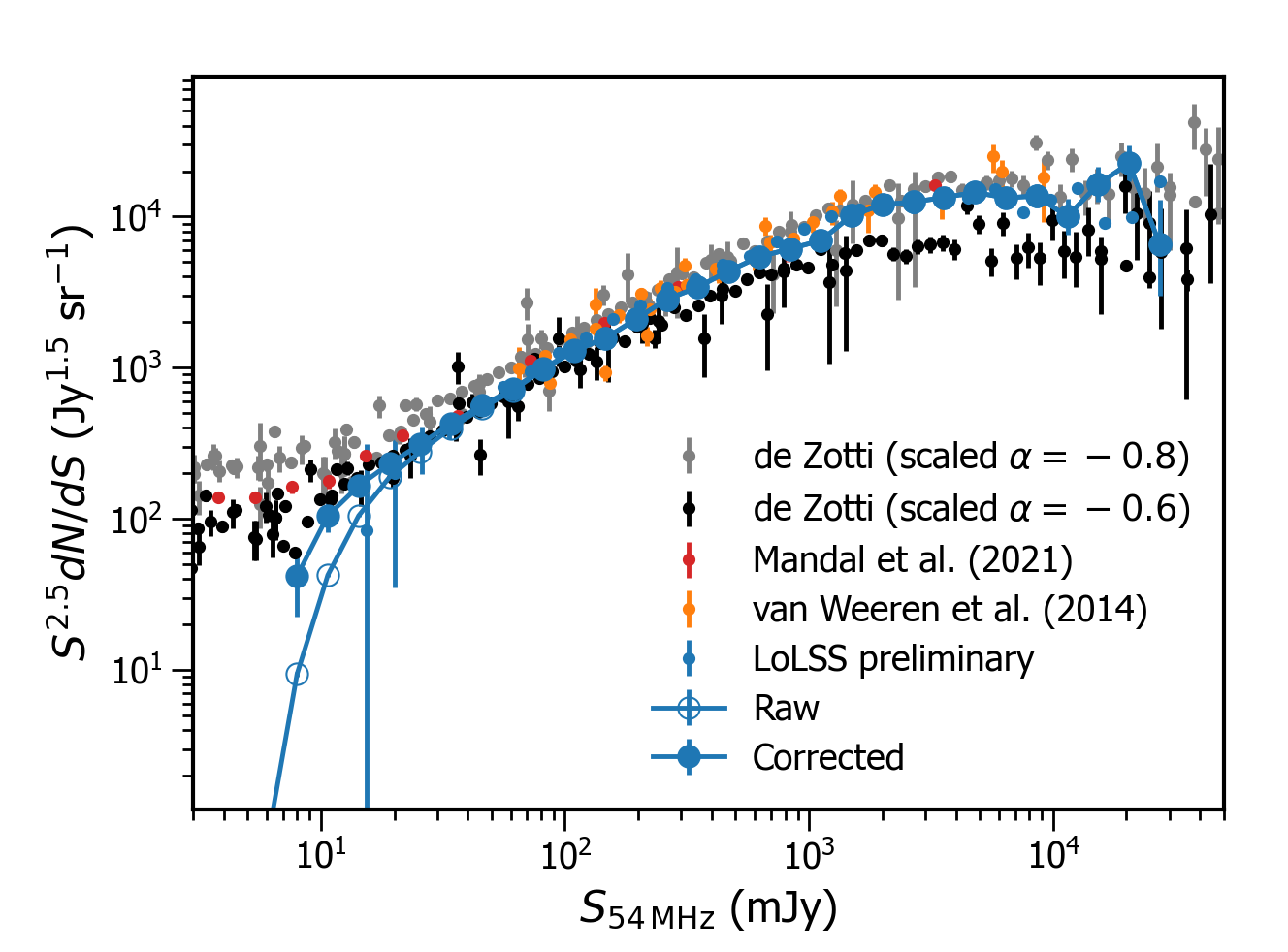}
 \caption{Euclidean-normalised differential source counts for \lol{} between 10\,mJy and 30\,Jy. The open circles show the raw, uncorrected source counts, while the filled circles show the counts corrected for completeness and resolution bias. The 1.4\,GHz source counts from various surveys compiled by \citet{deZotti2010}, and scaled to 54\,MHz  assuming a spectral index of -0.8 (in gray) and -0.6 (in black) are shown for comparison. Also for comparison are the LOFAR LBA source counts by \citet{vanWeeren2014} at $32$, $46$ and $62$ MHz (in orange) and the deep LOFAR $146$-MHz source counts by \citet{Mandal2021} (in red), all scaled to $54$ MHz with a spectral index of $-0.8$.  }
 \label{fig:sourcecount}
\end{figure}

The Euclidean-normalised differential source counts for \lol{} are plotted in Fig.~\ref{fig:sourcecount} where the uncertainties on the final normalized source counts were propagated from the error on the completeness correction and the Poisson errors \citep{Gehrels1986} on the raw counts per flux density bin. Incompleteness was accounted for by using the measured peak intensities to calculate the fractional area of the survey in which each source can be detected, $A_i$, with the total count in each flux density bin then determined as $N = \sum {1/A_{i}}$. The error on this correction was estimated from the measured uncertainty on each peak intensity and subsequent error on the visibility area. A small resolution bias correction,  taking into account the size distribution of sources, was made following \cite{Williams2016} and \cite{Prandoni2001}. The raw as well as corrected and normalised source counts in each flux density bin are listed in Table~\ref{tab:src_counts} along with the average visibility area correction (i.e. the average of the completeness correction) and the resolution bias correction in each bin.

The LoLSS source counts show good agreement with the counts derived from the preliminary release and reliably probe down to 10\,mJy compared to $\sim$25\,mJy for the preliminary release. They also agree well with the source counts derived by \cite{vanWeeren2014} from LOFAR LBA observations at 34, 46 and 62 MHz. As for the preliminary release, the source counts presented here show excellent agreement with higher frequency counts, with a change in the average spectral index of the population at lower flux densities. For comparison, we considered the  1.4\,GHz source counts compilation of \citet{deZotti2010}, scaled down to 54\,MHz assuming two different spectral indices. Above around 100\,mJy, the average spectral index is consistent with $-0.8$, while below this value a spectral index of $-0.6$ gives better agreement between the high and low frequency source counts. We also compare the LoLSS source counts to the very deep $146$-MHz source counts derived by \cite{Mandal2021} for the LOFAR Deep fields, which show a consistent change in average spectral index with flux density between $146$ MHz and $1.4$ GHz.  

\begin{table}[]
    \centering
    \begin{tabular}{ccccc}
    \hline
    $S$ & $<A_{vis}>$ & $R_{cor}$ & $dN$ & $S^{2.5}dN/dS$ \\
    ~[mJy] &           &           &      & [Jy$^{1.5}$sr$^{-1}$] \\
    \hline
$ 0.01$--$ 0.01$ & $ 0.47$ & $ 1.04 \pm  0.37$ & $  865^{ 30}_{ 29}$ & $42.00^{19.34}_{19.34}$ \\
$ 0.01$--$ 0.01$ & $ 0.66$ & $ 1.01 \pm  0.10$ & $ 2531^{ 51}_{ 50}$ & $105.05^{23.21}_{23.21}$ \\
$ 0.01$--$ 0.02$ & $ 0.80$ & $ 1.01 \pm  0.10$ & $ 4046^{ 65}_{ 64}$ & $165.03^{18.29}_{18.29}$ \\
$ 0.02$--$ 0.02$ & $ 0.89$ & $ 1.00 \pm  0.10$ & $ 4701^{ 70}_{ 69}$ & $231.60^{11.73}_{11.73}$ \\
$ 0.02$--$ 0.03$ & $ 0.94$ & $ 1.00 \pm  0.10$ & $ 4499^{ 68}_{ 67}$ & $312.38^{ 7.85}_{ 7.85}$ \\
$ 0.03$--$ 0.04$ & $ 0.96$ & $ 1.00 \pm  0.10$ & $ 4157^{ 65}_{ 64}$ & $429.19^{ 7.01}_{ 7.01}$ \\
$ 0.04$--$ 0.05$ & $ 0.98$ & $ 1.00 \pm  0.10$ & $ 3637^{ 61}_{ 60}$ & $562.73^{ 5.06}_{ 5.06}$ \\
$ 0.05$--$ 0.07$ & $ 0.99$ & $ 1.00 \pm  0.10$ & $ 3069^{ 56}_{ 55}$ & $723.53^{ 3.88}_{ 3.88}$ \\
$ 0.07$--$ 0.09$ & $ 0.99$ & $ 1.00 \pm  0.10$ & $ 2751^{ 53}_{ 52}$ & $989.60^{ 4.26}_{ 4.26}$ \\
$ 0.09$--$ 0.13$ & $ 0.99$ & $ 1.00 \pm  0.10$ & $ 2351^{ 49}_{ 48}$ & $1309.63^{ 3.01}_{ 3.01}$ \\
$ 0.13$--$ 0.17$ & $ 1.00$ & $ 1.00 \pm  0.10$ & $ 1835^{ 44}_{ 43}$ & $1567.46^{ 2.83}_{ 2.83}$ \\
$ 0.17$--$ 0.23$ & $ 1.00$ & $ 1.00 \pm  0.10$ & $ 1592^{ 41}_{ 40}$ & $2103.37^{ 3.97}_{ 3.97}$ \\
$ 0.23$--$ 0.30$ & $ 1.00$ & $ 1.00 \pm  0.10$ & $ 1395^{ 38}_{ 37}$ & $2839.28^{ 1.22}_{ 1.22}$ \\
$ 0.30$--$ 0.40$ & $ 1.00$ & $ 1.00 \pm  0.10$ & $ 1082^{ 34}_{ 33}$ & $3403.46^{ 0.52}_{ 0.52}$ \\
$ 0.40$--$ 0.54$ & $ 1.00$ & $ 1.00 \pm  0.10$ & $  897^{ 31}_{ 30}$ & $4362.36^{ 0.49}_{ 0.49}$ \\
$ 0.54$--$ 0.72$ & $ 1.00$ & $ 1.00 \pm  0.10$ & $  721^{ 28}_{ 27}$ & $5422.62^{ 0.75}_{ 0.75}$ \\
$ 0.72$--$ 0.97$ & $ 1.00$ & $ 1.00 \pm  0.10$ & $  527^{ 24}_{ 23}$ & $6129.83^{ 1.16}_{ 1.16}$ \\
$ 0.97$--$ 1.29$ & $ 1.00$ & $ 1.00 \pm  0.10$ & $  386^{ 21}_{ 20}$ & $6943.72^{ 1.80}_{ 1.80}$ \\
$ 1.29$--$ 1.73$ & $ 1.00$ & $ 1.00 \pm  0.10$ & $  366^{ 20}_{ 19}$ & $10182.51^{ 2.78}_{ 2.78}$ \\
$ 1.73$--$ 2.31$ & $ 1.00$ & $ 1.00 \pm  0.10$ & $  279^{ 18}_{ 17}$ & $12004.70^{ 4.30}_{ 4.30}$ \\
$ 2.31$--$ 3.10$ & $ 1.00$ & $ 1.00 \pm  0.10$ & $  188^{ 15}_{ 14}$ & $12510.63^{ 6.65}_{ 6.65}$ \\
$ 3.10$--$ 4.14$ & $ 1.00$ & $ 1.00 \pm  0.10$ & $  130^{ 12}_{ 11}$ & $13379.50^{10.29}_{10.29}$ \\
$ 4.14$--$ 5.54$ & $ 1.00$ & $ 1.00 \pm  0.10$ & $   91^{ 11}_{ 10}$ & $14484.81^{15.92}_{15.92}$ \\
$ 5.54$--$ 7.40$ & $ 1.00$ & $ 1.00 \pm  0.10$ & $   54^{  8}_{  7}$ & $13293.56^{24.62}_{24.62}$ \\
$ 7.40$--$ 9.90$ & $ 1.00$ & $ 1.00 \pm  0.10$ & $   36^{  7}_{  6}$ & $13706.50^{38.07}_{38.07}$ \\
$ 9.90$--$13.24$ & $ 1.00$ & $ 1.00 \pm  0.10$ & $   17^{  5}_{  4}$ & $10010.35^{58.88}_{58.88}$ \\
$13.24$--$17.71$ & $ 1.00$ & $ 1.00 \pm  0.10$ & $   18^{  5}_{  4}$ & $16392.66^{91.07}_{91.07}$ \\
$17.71$--$23.68$ & $ 1.00$ & $ 1.00 \pm  0.10$ & $   16^{  5}_{  4}$ & $22535.84^{140.85}_{140.85}$ \\
$23.68$--$31.68$ & $ 1.00$ & $ 1.00 \pm  0.10$ & $    3^{  3}_{  2}$ & $6535.09^{217.84}_{217.84}$ \\
\hline 
    \end{tabular}
    \caption{Euclidean-normalised differential source counts for \lol{} between 10\,mJy and 30\,Jy}
    \label{tab:src_counts}
\end{table}

\subsection{Preliminary analysis of LoLSS Stokes V data}

As detailed in Sec.~\ref{sec:pipelinedd}, Stokes\,V mosaics are produced by the LoLSS data reduction pipeline. Pulsars, brown dwarfs, stars, and exoplanets are known to emit highly circularly-polarised radiation \citep[e.g.][]{Zarka1998, Vedantham2020, Callingham2021, Turner2021}.

We searched the Stokes\,V mosaics of LoLSS for any significant circularly-polarised sources. We find the leakage of total intensity into Stokes\,V to be $\approx$0.5\%, a factor of five worse than that found for V-LoTSS (Callingham et al. in prep). Such a difference between LoLSS and LoTSS is likely due to larger Faraday rotation corrections performed at lower-frequencies and possible mechanical issues with the LBA dipoles.

\begin{figure}
\centering
 \includegraphics[width=.5\textwidth]{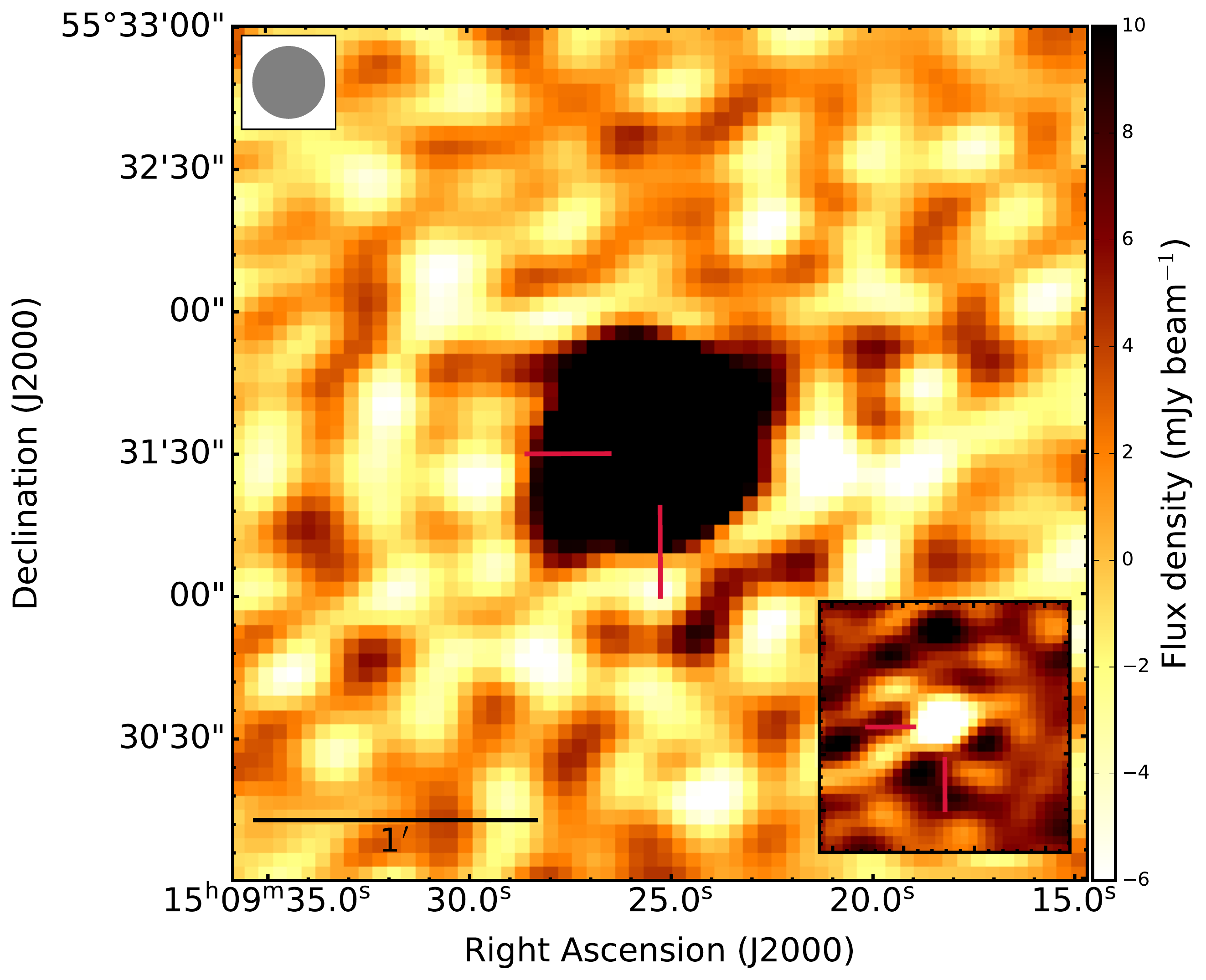}
 \caption{Total intensity image of PSR\,B1508+55 with a Stokes\,V image inset. The red cross-hairs point to PSR\,B1508+55, and the synthesised beam is displayed at the top-left corner. The total time and frequency average flux density of PSR\,B1508+55 in Stokes\,I and V is 1.42\,Jy and -25\,mJy, respectively.}
 \label{fig:stokesv}
\end{figure}

For this release of LoLSS we find only one circularly-polarised source that is not consistent with leakage. The source is pulsar PSR\,B1508+55, as shown in Figure\,\ref{fig:stokesv}, and has a circularly-polarised fraction of 1.6$\pm$0.3\% and is an 18$\sigma$ detection in Stokes\,V. A complete analysis of the LoLSS Stokes\,V properties will be the focus of a follow-up manuscript but the detection of PSR\,B1508+55 demonstrates that LoLSS Stokes\,V products are science-ready.

\section{Summary}
\label{sec:summary}

In this work, we have presented the first data release of the LOFAR LBA Sky Survey (LoLSS). The interferometer visibilities were automatically processed using the Pipeline for LOFAR LBA (PiLL); compared with the preliminary release, we have now included the derivation and correction of direction-dependent systematic errors. The typical rms noise of the survey is 1.55~\mjybeam{} at a resolution of 15\arcsec, although several regions of the survey are limited in sensitivity by dynamic range issues that affect the surroundings of sources with flux density $S_\nu > 100$~mJy.

% survey details
The footprint of the survey is about 650 deg$^2$ and the final catalogue contains 42\,463 sources. The catalogue is 50\% complete at 5.2 mJy and 95\% complete at 11 mJy. The fraction of false positives is estimated at 1.47\%. Compared to the LOFAR Two-metre Sky Survey (LoTSS), where most (92\%) of the detected sources are point-like, in LoLSS this fraction decreases to 82\% despite the lower resolution. This is likely due to the lower sensitivity, which favours the detection of nearby sources. The survey astrometric precision is estimated to be $\sigma_{\rm RA} = 1.48\arcsec$ and $\sigma_{\rm Dec} = 1.17\arcsec$. The flux scale accuracy is estimated to be 6\%. This has been validated by cross-checking LoLSS flux densities with the values derived from existing surveys. 

Six in-band flux density measurements are also derived (44, 48, 52, 56, 60, and 64 MHz). While their median values are in line with the full-band catalogue, the in-band spectral index appears systematically flatter (about $+0.2 - +0.3$) compared to expectations. However, this systematic offset cannot account for the full flattening, which is interpreted to be partially physical.

% future prospects (sun cycle, lofar 2, other data)
We plan to complete the observations of the northern sky at Dec~$>24\deg$ by the end of 2022. These observations and those taken in the past years will be included in a forthcoming release of LoLSS. Starting from 2024, LOFAR will enter into a phase of renovation leading to ``LOFAR 2.0'' (Hessels et al. in prep.). This process will enable simultaneous LBA+HBA observations for better ionospheric calibration, the use of all LBA dipoles increasing the collecting area of each station, and the possibility to exploit a larger number of beams increasing the survey speed. Finally, a large superstation which has been completed in France \citep[NenuFAR;][]{Zarka2012} will boost LOFAR-VLBI sensitivity in the LBA band and potentially permit the use of international baselines in future surveys.

\begin{acknowledgements}

% LOFAR
LOFAR is the LOw Frequency ARray designed and constructed by ASTRON. It has observing, data processing, and data storage facilities in several countries, which are owned by various parties (each with their own funding sources), and are collectively operated by the ILT foundation under a joint scientific policy. The ILT resources have benefited from the following recent major funding sources: CNRS-INSU, Observatoire de Paris and Universit\'e d’Orleans, France; BMBF, MIWF-NRW, MPG, Germany; Science Foundation Ireland
(SFI), Department of Business, Enterprise and Innovation
(DBEI), Ireland; NWO, The Netherlands; The Science and
Technology Facilities Council, UK; Ministry of Science and Higher Education, Poland; Istituto Nazionale di Astrofisica (INAF). This research has made use of the University of Hertfordshire high-performance computing facility (\url{https://uhhpc.herts.ac.uk/}) and the LOFAR-UK compute facility, located at the University of Hertfordshire and supported by STFC [ST/P000096/1].

% personal
FdG and MB acknowledge support from the Deutsche Forschungsgemeinschaft under Germany's Excellence Strategy - EXC 2121 “Quantum Universe” - 390833306.
FdG and GB acknowledge support from INAF PRIN MAINSTREAM "Galaxy cluster science with LOFAR".
MJH acknowledges support from the UK Science and Technology Facilities Council (ST/R000905/1).
RJvW acknowledges support from the ERC Starting Grant ClusterWeb 804208.
ABon and ABot acknowledge support from ERC Stg DRANOEL n. 714245. ABon acknowledges support from MIUR FARE grant “SMS”.
DJB acknowledges funding from the German Science Foundation DFG, via the Collaborative Research Center SFB1491 “Cosmic Interacting Matters - From Source to Signal.
IP acknowledges support from INAF under the SKA/CTA PRIN “FORECaST” and the PRIN MAIN STREAM “SAuROS” projects.
M.S.S.L. Oei acknowledges support from the VIDI research programme with project number 639.042.729, which is financed by The Netherlands Organisation for Scientific Research (NWO).
VC acknowledges support from the Alexander von Humboldt Foundation.
MK acknowledges support from the German Science Foundation DFG, via the Research Unit FOR 5195 "Relativistic Jets in Active Galaxies".
\end{acknowledgements}

\bibliographystyle{aa}
\bibliography{library}

\end{document}